\documentclass{emulateapj}

\usepackage{amsmath,amssymb,gensymb,times,graphics,morefloats,color}
\usepackage{afterpage, bm}
\usepackage{natbib}
\def\mearth{M\mkern-2mu Earth}
\def\kepler{\emph{Kepler}}

\def\msun{M_{\odot}}

\def\rsun{R_{\odot}}
\def\teff{T_\mathrm{eff}}
\def\feh{\mathrm{[Fe/H]}}

\def\K{$K$}

\def\vsini{$v\sin{i}$}

\shorttitle{The rotation and kinematics of nearby M dwarfs}
\shortauthors{Newton et al.}

\begin{document}

\title{The rotation and Galactic kinematics of mid M dwarfs in the solar neighborhood}

\author{Elisabeth R. Newton\altaffilmark{1}, Jonathan Irwin\altaffilmark{1}, David Charbonneau\altaffilmark{1}, Zachory K. Berta-Thompson\altaffilmark{2}, Jason A. Dittmann\altaffilmark{1}, Andrew A. West\altaffilmark{3}}
\altaffiltext{1}{Harvard-Smithsonian Center for Astrophysics, 60 Garden Street, Cambridge, MA 02138, USA}
\altaffiltext{2}{Kavli Institute for Astrophysics and Space Research and Department of Physics, Massachusetts Institute of Technology, Cambridge, MA 02139, USA}
\altaffiltext{3}{Department of Astronomy, Boston University, 725 Commonwealth Avenue, Boston, MA 02215, USA}

\begin{abstract}
Rotation is a directly-observable stellar property, and drives magnetic field generation and activity through a magnetic dynamo. Main sequence stars with masses below approximately $0.35\msun$ (mid-to-late M dwarfs) are fully-convective, and are expected to have a different type of dynamo mechanism than solar-type stars.  Measurements of their rotation rates provide insights into these mechanisms, but few rotation periods are available for these stars at field ages. Using photometry from the MEarth Project, we measure rotation periods for $387$ nearby, mid-to-late M dwarfs in the Northern hemisphere, finding periods from $0.1$ to $140$ days. The typical rotator has stable, sinusoidal photometric modulations at a semi-amplitude of $0.5$ to $1\%$. We find no period-amplitude relation for stars below $0.25\ \msun$ and an anti-correlation between period and amplitude for higher-mass M dwarfs. We highlight the existence of older, slowly-rotating stars without H$\alpha$ emission that nevertheless have strong photometric variability. We use parallaxes, proper motions, radial velocities, photometry and near-infrared metallicity estimates to further characterize the population of rotators. The Galactic kinematics of our sample is consistent with the local population of G and K dwarfs, and rotators have metallicities characteristic of the Solar Neighborhood. We use the $W$ space velocities and established age-velocity relations to estimate that stars with $P<10$ days have ages of on average $<2$ Gyrs, and that those with $P>70$ days have ages of about $5$ Gyrs. The period distribution is mass dependent: as the mass decreases, the slowest rotators at a given mass have longer periods, and the fastest rotators have shorter periods. We find a lack of stars with intermediate rotation periods, and the gap between the fast and slow rotating populations is larger for lower masses. Our data are consistent with a scenario in which these stars maintain rapid rotation for several Gyr, then spin down rapidly, reaching periods of around $100$ days by a typical age of $5$ Gyr.

\end{abstract}


\section{Introduction}

Rotation is one of the few directly observable stellar properties, and these observations provide constraints on the angular momentum evolution of stars. Late-time angular momentum loss is governed by magnetized stellar winds, which depend on the magnetic field topology. Stellar rotation therefore provides empirical insight into both the stellar wind and magnetic field. These interlinked properties are also relevant to the detection and characterization of their planetary systems. Stellar winds and magnetic fields affect habitability, potentially stripping the planetary atmosphere \citep{Cohen2014}. Rapidly-rotating, active stars are also difficult targets for exoplanet surveys. Line broadening from the most rapidly-rotating stars and radial velocity signals induced by stellar activity pose challenges for radial velocity surveys \citep[][]{Saar1997,Saar1998}.  In transit surveys, periodic modulations from star spots can cause confounding signals \citep[e.g.][]{Berta2012}.

The rotation of fully convective stars \citep[with $M<0.35\ \msun$;][]{Chabrier1997} and of those that have an outer convective envelope is both age- and mass-dependent. Stars spin up as they approach the zero-age main sequence, a consequence of Kelvin-Helmholtz contraction and late-stage accretion, and the conservation of angular momentum. To match the rotation periods observed in the youngest clusters, pre-main sequence stars must also experience early angular momentum losses \citep[e.g.][]{Hartmann1989,Bouvier1997}. This is thought to be the result of star-disk interactions \citep[e.g.][]{Koenigl1991,CollierCameron1995,Matt2005}. 
After reaching the main sequence, angular momentum loss is dominated by magnetized stellar winds, the strength of which may depend on mass. By the age of the Hyades, the rotation periods of solar-type dwarfs have reached a narrow range and subsequently obey a Skumanich-type relation \citep{Skumanich1972} between angular velocity ($\omega$), mass, and age ($t$), where $\omega \propto t^{-1/2}$. The well-defined rotation-age relation and the lack of dependence on initial conditions gives rise to the concept of gyrochronology \citep{Barnes2003}. 

The convergence of stellar rotation periods can be described by a parametrized wind-loss model \citep{Kawaler1988, Reiners2012b}, in which more rapid rotators spin down faster.
The rate of angular momentum loss is thought to saturate for stars with angular velocities faster than some mass-dependent critical value $\omega_{\rm sat}$. This is typically modeled as a change in how the angular momentum loss rate depends on rotation rate, which occurs when the angular velocity drops below the critical value. This leads to a change in the time dependence of the rotation rate itself. In the most common prescription \citep[e.g.][]{Bouvier1997}, during the saturated regime (rapid rotation), $\omega$ decreases exponentially with time, and in the unsaturated regime, $\omega$ follows the Skumanich law. This behavior causes the rotation periods to converge as the stars evolve from the saturated to unsaturated regime.
The well-behaved relationship between rotation, age, and color for solar-type stars with ages older than roughly $650$ Myr can be used to infer ages of isolated field stars through gyrochronology, by measuring their rotation period and color. 

For solar-type stars, angular momentum evolution can be modeled with reasonable success using presently-available observations. 
However, these models may not be able to simultaneously fit the evolution of the lowest mass dwarfs \citep[e.g.][]{Irwin2011,Reiners2012b}. This could arise from a different magnetic dynamo in fully-convective stars. Models of the solar magnetic dynamo indicate the importance of stellar rotation in the generation of the solar magnetic field \citep[see][for a review]{Charbonneau2005}. The tachocline, the interface between the radiative and convective zone, has also been thought to be key. In some solar dynamo models, the tachocline is where the strengthening and organization of the solar magnetic field occurs. The tachocline is not present in fully-convective stars; nevertheless, strong magnetic fields appear to be prevalent amongst these low-mass stars \citep{Johns-Krull1996, Reiners2010}. Theoretical models of magnetic dynamos in fully-convective stars focus on how rotation and convection can maintain a magnetic field in the absence of a tachocline \citep[e.g.][]{Chabrier2006,Dobler2006, Browning2008,Yadav2015}.

Measurements of stellar rotation provide insight into angular momentum evolution and the magnetic dynamo. Observational constraints at young ages come predominantly from open clusters and moving groups, with ages of a few Myr (e.g. the Orion Nebular Cluster) to the Hyades and Praesepe \citep[$650$ Myr;][]{Perryman1998,Gaspar2009,Bell2014}.
The Sun and the old stellar clusters NGC 6811 ($1$ Gyr) and NGC 6819 ($2.5$ Gyr) provide anchors at older ages for stars around solar mass \citep[]{Meibom2011, Meibom2015}. 
While solar-mass stars have converged to a narrowly-constrained mass-rotation relation by $650$ Myr, M dwarfs still show a broad range of rotation rates at this age and continue to undergo substantial angular momentum evolution at field ages. Observations of \emph{field} M dwarfs are therefore particularly important for constraining their angular momentum evolution.
Substantial observational progress has been made in recent years, with many new measurements of rotation periods for field M dwarfs, notably by \citet[]{Kiraga2007}, \citet{Norton2007}, \citet{Hartman2011}, \citet{Irwin2011},  \citet{Kiraga2012}, \citet{Goulding2013}, \citet{Kiraga2013}, and \citet{McQuillan2013}.

Determining the ages of field stars is important for enabling their use in modeling rotational evolution. M dwarf radii and effective temperatures remain mostly unchanged once they reach the main sequence and therefore do not provide robust constraints on their ages. Galactic kinematics provide one possible avenue: The motions of stars through the solar neighborhood bear signatures of their past dynamical interactions.
Stars form in dense clusters with kinematics generally constrained to a co-rotating disk with a small scale height and a small velocity dispersion. After their formation, there are two primary effects on the stars' kinematics: the clusters dissipate, and stars undergo dynamical heating. Most clusters are not gravitationally bound and evaporate over time, although some physically-associated, coeval stellar populations persist as the young moving groups and open clusters mentioned previously. Dynamical interactions increase the velocity dispersion of a group of stars over time. This mechanism acts within the kinematically cold stellar population known as the ``thin disk,'' and also is thought to have produced the population of kinematically hotter stars often referred to as the ``thick disk''. Whether the thick disk should be described by a single population or a superposition of many mono-age or mono-abundance populations has recently been called into question \citep{Bovy2012, Minchev2015}, but it is composed of stars generally older than, and with different chemical abundances from, the canonical thin disk \citep[][]{Bensby2005}. 
Within the thin/young disk, the velocity dispersion of a group of stars increases with time. Relationships between age and velocity dispersion have calibrated for stars in the Solar Neighborhood \citep[e.g.][]{Wielen1977,Nordstrom2004}. Kinematics can therefore shed light on the ages of \emph{populations} of stars. 

\citet{Irwin2011} contributed a substantial number of the currently-available measurements for fully-convective stars, with rotation periods for $41$ M dwarfs from the MEarth transit survey \citep{Berta2012, Irwin2014} that had published trigonometric parallaxes. By assigning stars to the thin/young and thick/old disk based on their space velocities, we showed that the rapidly-rotating M dwarfs were on average younger than the slowly-rotating stars. 

In this work, we extend the analysis of \citet{Irwin2011} to the full Northern sample of M dwarfs observed by MEarth. Our sample is particularly of interest due to the large body of observations that our team has gathered on these stars. We measured parallaxes for $1507$ of the MEarth targets using MEarth astrometry \citep{Dittmann2014} and calibrated the MEarth photometric bandpass to provide optical magnitudes for every target \citep{Dittmann2015}. In \cite{Newton2014}, we obtained low-resolution near-infrared spectra of $447$ MEarth targets, measuring their absolute radial velocities and estimating their iron abundances ($\feh$). Using the H$\alpha$ line to trace magnetic activity and additional rotation periods derived from MEarth data, we found that the fraction of active stars continues to decrease with increasing rotation period out to the longest rotation periods in the MEarth sample \citep{West2015}.

\section{Photometry from MEarth}

The MEarth Project is an all-sky transit survey searching for planets around approximately 3000 nearby, mid-to-late M dwarfs \citep{Berta2012, Irwin2014}. MEarth-North is located at the Fred Lawrence Whipple Observatory, on Mount Hopkins, Arizona, and has been operational since 2008 September. MEarth-South, at Cerro Tololo Inter-American Observatory (CTIO) in Chile, was commissioned in January 2014. Each installation consists of eight 40cm telescopes on German Equatorial Mounts, equipped with CCD cameras. This work uses data from only MEarth-North.

\citet{Nutzman2008} selected the Northern MEarth targets from the \citet{Lepine2005a} Northern proper motion catalog, which includes stars with proper motions $>0\farcs15\ \mathrm{yr}^{-1}$. MEarth exclusively targets nearby, mid-to-late M dwarfs: at the time of selection, all stars had parallaxes or distance estimates \citep[spectroscopic or photometric;][]{Lepine2005} placing them within 33 pc, and estimated stellar radii less than $0.33\ \rsun$. Trigonometric parallaxes are now available for the majority of stars in the Northern sample \citep{Dittmann2014}.

MEarth targets are spread across the sky and must therefore be targeted individually; targets are visited at a cadence of 20--30 minutes for observations taken as part of the main planetary transit search. Additional observations of all targets have been taken at a cadence of approximately 10 days since September 2011 for the astrometric program \citep{Dittmann2014,Dittmann2015}. We also include these data in the analysis presented here. Exposure times are set for each object such that a 2 Earth-radius planet transit would be detected in each datum at $3\sigma$, and therefore depend on the estimated stellar radii. We use the \citet{Delfosse2000} mass-$\mathrm{M}_K$ relation to estimate stellar masses, then the \citet{Bayless2006} mass-radius polynomial to estimate stellar radii. Our current exposure time calculations also use the trigonometric distances from \citet{Dittmann2014}. We did not adjust the exposure time of individual images when we updated the stellar parameters in order to avoid changing the effect of non-linearity on our photometry; instead, we co-add exposures to reach the requisite sensitivity when necessary. Each visit to a star may therefore contain multiple exposures.

Northern target stars are typically observed at the 20--30 minute cadence for one to two observing seasons, with each season lasting from mid-September of one year to mid-July of the following year. During southern Arizona's summer monsoons, MEarth-North is shut down completely. 

For the 2008--2009 and 2009--2010 observing seasons, we used long-pass filters composed of $5$ mm thick Schott RG715 glass. In 2010--2011, in an attempt to mitigate color-dependent systematics (discussed at the close of this section and in \citealt{Irwin2006}), we switched to a custom filter with a sharp interference cutoff at the red end, approximating the Cousins I-band. Finding that this increased the level of systematics rather than mitigating them, from 2011--2012 onwards we reverted to RG715 filters, but with $3$ mm thickness.

We do not attempt to stitch observations taken with different filters or different telescopes together, so the data on each star may be composed of multiple light curves, where we define a \emph{light curve} as the set of observations from a single MEarth telescope with a single filter setup. A single light curve will therefore contain data from one of the 2008--2010, 2010--2011, or 2011--2015 intervals.  Each object is usually assigned to a single telescope for the entirety of its observations; however, starting in the 2012--2013 season, two telescopes were assigned to a subset of the targets \cite[see][]{Berta2012}.  A small number of targets also appear in multiple fields (where there are multiple targets within the field of view) so may have more light curves.

For our data reduction, we follow the methodology of \citet{Irwin2006}, modified for the specifics of the MEarth data as detailed in \citet{Berta2012}. We highlight here several systematics that affect our ability to detect rotation periods:
\begin{enumerate}
\item{\cite{Irwin2011} noted weather-dependent effects in the differential magnitudes of the target M dwarfs, which result from variations in telluric water vapor absorption in the bandpass of the RG715 filter, or humidity dependence of the interference cutoff in the interference filters used in 2010--2011. Because our targets are typically the only M dwarfs in the field, the reference stars predominantly have bluer colors. Therefore, the observed fluxes of the targets and reference stars are affected differently by the varying telluric absorption or humidity when integrated over the filter bandpass. This effect cannot be corrected with standard differential photometry procedures, and we note that the resulting systematic effects are dominated by the time variability of the driving quantity (precipitable water vapor or humidity) and are not strongly correlated with airmass, so cannot be corrected by standard methods for removing atmospheric extinction. Instead, the differential magnitudes of all of the M dwarfs being observed within a half-hour window are combined to produce a lower cadence comparison light curve, which we call the ``common mode,'' that measures any residual photometric variations that are common to the target M dwarfs. Due to differences in the target spectral types, it is necessary to scale the common mode by a factor that varies for each object. This scale factor has proved difficult to calculate, so it is fit empirically from the light curve.}
\item{The MEarth telescopes use German Equatorial Mounts, which require the telescope tube to be flipped over the pier when the target crosses the meridian, resulting in rotation of the focal plane relative to the sky by $180\degree$, and causing stars to sample two distinct regions of the detector. Residual flat-fielding errors result in offsets in the differential magnitudes between the two locations. To correct for this, we assign a different baseline magnitude to observations taken at these two rotation angles. Additional flat fielding errors are inevitably introduced whenever the cameras are removed from the telescopes for repair, so we introduce a new pair of baseline magnitudes each time this is done. We refer to the set of data taken between these camera removals on a single side of the meridian as a light curve ``segment'', where each segment is modeled with its own baseline magnitude when producing differential photometry.}
\item{The large-scale structure of our flat field evolves on timescales of several years. We take nightly twilight flats, but because the illumination is dominated by scattered light, we filter out all of the large-scale structure. To account for the large-scale flat field structure, we observe a star field in the Galactic plane dithered randomly inside a one square degree box, and use photometry to obtain the large scale flat field pattern. These observations require a substantial amount of telescope time during photometric conditions, so are repeated only intermittently. We have used a single large-scale correction for each of the 2008--2010, 2010--2011, and 2011--2015 data sets, in order to avoid introducing spurious signals when there are sudden changes in this correction. However, there are several instances where rapid evolution in the flat-field is evident, which we account for by introducing additional segments with new baseline magnitudes.}
\end{enumerate}
We discuss our treatment of these systematics during rotation period determination in the following section.

\section{Determining rotation periods}\label{Sec:method}

We attempt to identify rotation periods in each of the $1883$ targets observed with MEarth-North, including data obtained through 16 August 2015.

\subsection{Period detection}\label{Sec:method_period}

We apply the method used by \citet{Irwin2011}, which is based on \citet{Irwin2006}, to fit simultaneously for terms needed to account for both our systematics, and for rotational modulation. For each light curve, we fit both a null hypothesis, which assumes that the light curve has no astrophysical variability and can be fit with systematics alone, and an alternate hypothesis that includes a sinusoid. 

Our models include two terms to address the systematics discussed in the previous section: the common mode, and the baseline magnitude in each segment of the light curve. 

The null model takes the form:
\begin{align}
m_\mathrm{null}(t) =  m_i + k\: c(t) 
\end{align}
where $i$ is the segment number, and $m_i$ is the baseline magnitude
for light curve segment $i$, $t$ is time and $k$ scales the common
mode $c(t)$.  We only include as many $m_i$ constants as there are
segments containing data points, so for example if a target was
observed only on one side of the meridian, only a single baseline
magnitude is fit. This model corresponds to a constant intrinsic
magnitude (above the Earth's atmosphere) for the target M dwarf,
modulated by the atmospheric and instrumental systematics.

The alternate model additionally includes a sinusoid:
\begin{align}\label{Eq:sine}
m_\mathrm{alt}(t) = m_\mathrm{null}(t) +  a\: \sin(\omega t + \phi)
\end{align}
where $a$ is the semi-amplitude (in magnitudes), $\phi$ is the phase, and $\omega$ is the angular frequency $\omega= 2\pi/P$, where the rotation period is given by $P$. For fitting purposes, we rewrite the sine term on the right-hand side of this equation using standard trigonometric identities to replace the non-linear $\phi$ parameter with a pair of linear semi-amplitudes $a_s$ and $a_c$:
\begin{align}
m_\mathrm{alt}(t) = m_\mathrm{null}(t) + a_s\: \sin(\omega t) + a_c\: \cos(\omega t)
\label{malt}
\end{align}

Observations of a star may comprise several separate light curves. These are fit simultaneously, enforcing a common period over all light curves. The common mode scaling, baseline magnitudes, and the amplitude and phase of the sinusoid are independent. Before fitting, we remove data deviating from the median by more than $5\sigma$, where we use the median absolute deviation scaled to the Gaussian-equivalent RMS \citep{Hoaglin1983} to define $\sigma$. Clipping is done to remove flares and (in some cases) eclipses, rather than to iteratively improve our fit by removing outliers. We do not use outlier removal in our fits because we are comparing models at different periods -- each model could clip different data, and the $\chi^2$ of poorly-fitting models would be artificially reduced. 
 
We use a maximum likelihood method to find the best-fitting rotation period under the assumption of the alternate hypothesis. We step through a uniformly spaced grid of frequencies corresponding to periods ranging from $0.1$ to $1500$ days, performing a linear least-squares fit of Eq. (\ref{malt}) to the remaining variables at each frequency. As the null hypothesis is nested within the alternate, an F-test is appropriate for determining whether the addition of a sinusoid is warranted. We therefore calculate the F-test statistic (which measures the amount of variance that is explained by the additional parameters in the alternate model) at each frequency and select the one with the highest statistic as the candidate frequency. The set of F-test statistic values as a function of frequency are analogous to a periodogram.

We then visually inspect the light curves for each M dwarf, looking at the data with the common mode and varying baseline magnitudes removed. We look at differential magnitude as a function of time and at the data phased to the candidate frequency from the F-test. For some objects, multiple exposures were acquired at each visit in order to achieve the S/N we require for planet detection. While fitting is performed on un-binned data, we  visually examine the data both binned by visit and un-binned.

We assess the validity of the candidate period by posing a series of questions developed after early exploration of the data, but emphasize that the criteria we use in deciding whether a period is detected are fundamentally qualitative. The questions we ask are:
\begin{itemize}
\item{Can the candidate periodic rotation signal be seen by eye in the binned, phase-folded data?}
\item{Are two or more complete, near-consecutive rotation cycles seen? An important factor are the baseline magnitudes, which can allow data at disparate times to be offset arbitrarily; thus, data spanning multiple segments must be considered carefully.}
\item{Is the candidate rotation signal uncorrelated with systematics included in the model (baseline magnitude offsets and the common mode) and with the FWHM of the image?}
\item{If the candidate period is $<10$ days, can the variability be seen during single, well-sampled nights?}
\item{If there are simultaneously-observed light curves, do the concurrent data agree?}
\end{itemize}
After considering these questions, we classify objects as either ``rotators'' or ``non-rotators''. Rotators are objects that we consider to have secure detections of periodic photometric modulation that we assume to be attributable to stellar rotation. We further assign rotators a rating of ``grade A'' ($274$ stars) or ``grade B'' ($113$ stars). Grade A means that we are confident that we have identified a sinusoidal photometric modulation that can be attributed to an astrophysical source; the answer to all posed questions must be ``yes''. Grade B means that a modulation has been detected that we believe to be real, but that the signal does not pass all of our tests. Most grade B rotators fail only one criterion, and fall into one of the following categories: 1) two complete cycles are not seen, but the variability that is detected strongly suggests periodic modulation, 2) the only data available are from our astrometric program so the candidate periodicity is not sampled at high-cadence, 3) a convincing period is detected, but the noise level is comparable to amplitude of modulation.

\begin{figure*}
\begin{center}
\begin{tabular}{ccc}
\includegraphics[width=\linewidth]{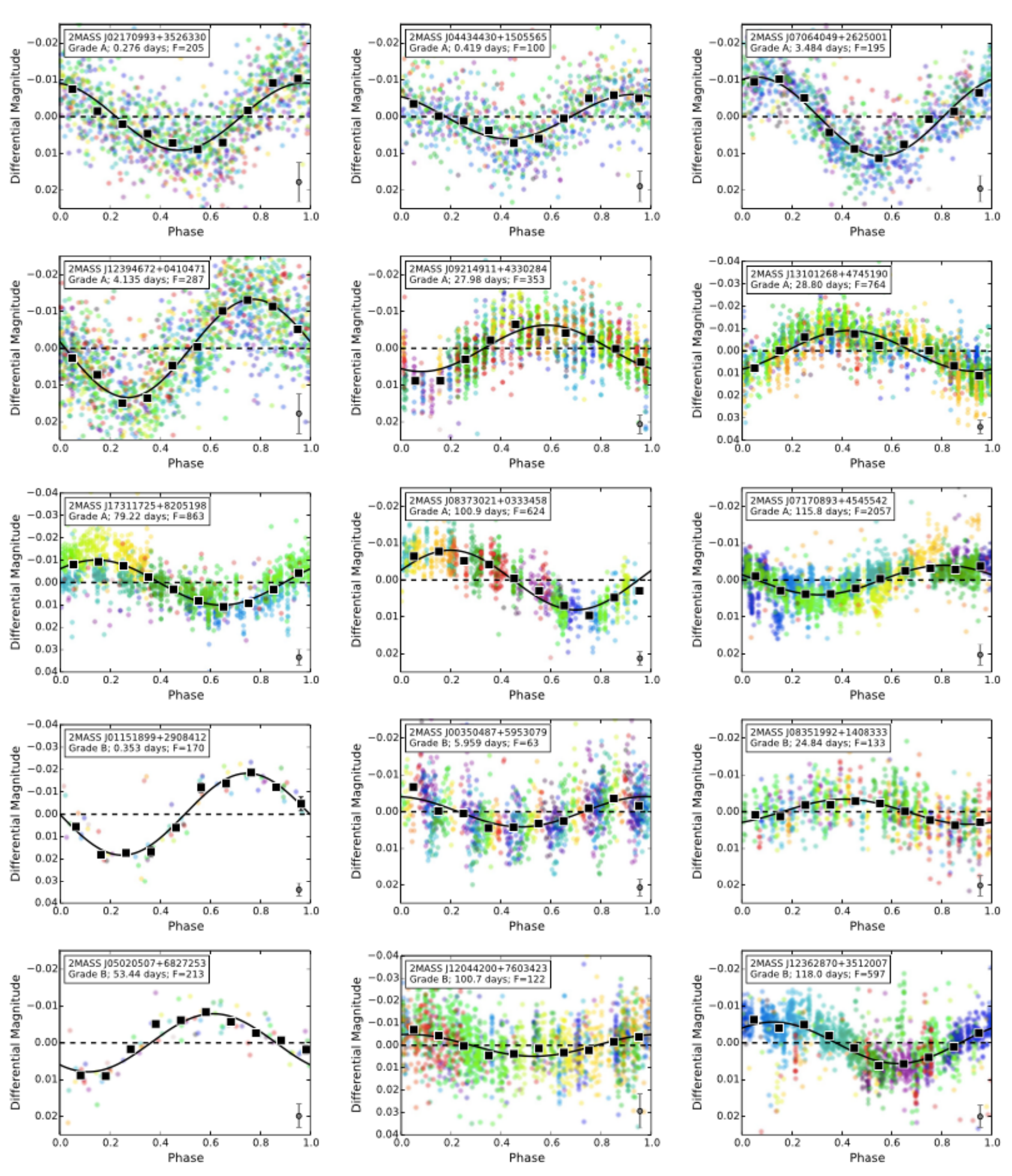}
\end{tabular}
\caption{Examples of typical rotators, randomly selected from our sample. The top three rows show grade A rotators, and the bottom two rows show grade B rotators. Data points are colored according to the observation number, and the median error is indicated in the bottom right corner. The earliest data points are purple, the latest are red. We also show the median magnitude in ten uniformly spaced bins in phase; the sample scatter about the median is plotted but typically smaller than the plotting symbol. The label indicates the rating (grade A or B), the rotation period, and the F-test statistic (F).
\label{Fig:examples}}
\end{center}
\end{figure*}

Representative examples of grade A and B rotators are shown in Fig. \ref{Fig:examples}. The median of the phased data in ten equally-spaced bins is also included. The sample scatter about the median ($1.48$ times the median absolute deviation) is shown, but is typically smaller than the data points.

We assign non-rotators a rating of either ``possible/uncertain'' ($239$ stars) or ``no detection'' ($1260$ stars). If we detect a candidate signal, but are not confident enough in its veracity to consider the object a rotator, we assign it a ``possible'' rating. A ``no detection'' rating indicates that we cannot positively identify a periodic modulation. Importantly, ``no detection'' does not mean that object is not rotating. 

\begin{figure}
\includegraphics[width=\linewidth]{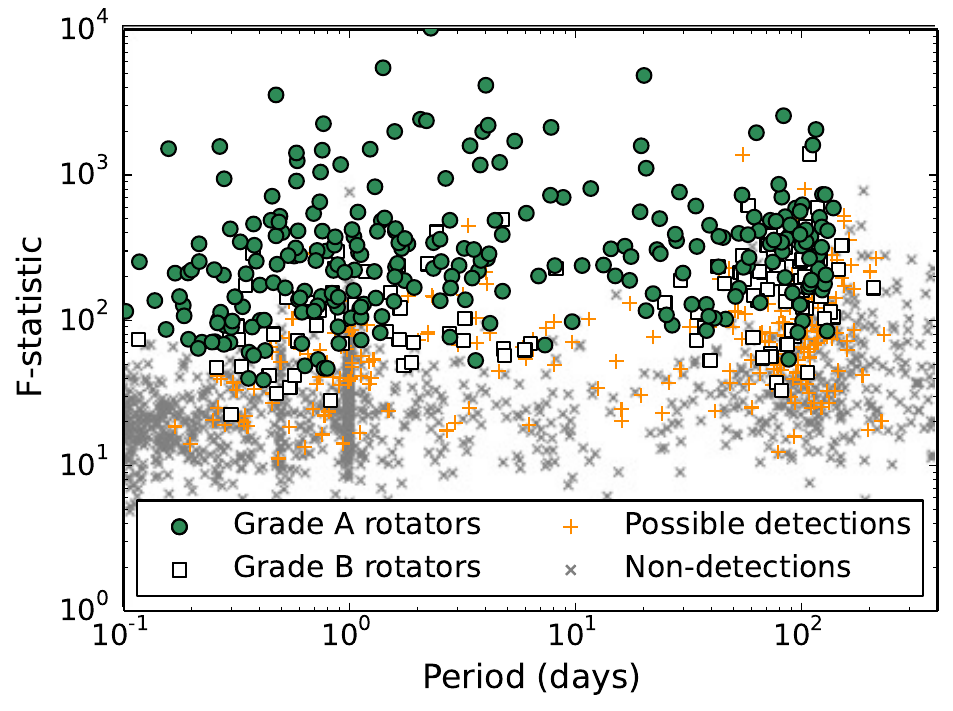}
\caption{F-test statistic as a function of rotation period for grade A (filled circles) and grade B (open squares) rotators, ``possible'' detections (plus symbols), and non-detections (crosses). 
\label{Fig:ftest}}
\end{figure}

While we do not require a specific value for the F-test statistic, our rotators comprise most of the statistically-significant peaks (Fig. \ref{Fig:ftest}).

We present rotation period measurements for rotators with grade A and B ratings in a machine-readable table available in the online version of this article. We do not attempt to assign errors in these periods (for example, based on the width of the periodogram peak) because there are usually multiple peaks in the periodogram, and an estimate based only on the dispersion about one particular peak would be misleading. We refer the reader to \citet{Irwin2011} for details of signal injection and recovery tests which can be used to gauge approximate period errors. 

We include estimates of stellar mass and radius. Our stellar masses are estimated from the absolute $K$ magnitude, using the relation from \citet{Delfosse2000}, which we modified to allow extrapolation past the limits of the calibration. The relation is unphysical beyond the calibration range of $4.5<M_K<9.5$, and a number of our stars are fainter than this limit. Our modification simply enforces a constant value and first derivative slope at the boundaries, and produces a physically reasonable result. For stellar radii, we use the mass-radius relation from \citet{Boyajian2012}.

We also include non-rotators with ``possible'' or ``no detection'' ratings in the online-only table, listing the period of the strongest peak in the periodogram. We caution that these periods should not be interpreted as detections. Additional data would be useful for confirming or disproving the listed periods.

The rotation periods and ratings in this paper supersede those reported previously in \citet{Irwin2011} and \cite{West2015}.

\subsection{Identifying multiples}

The multiplicity fraction amongst mid M dwarfs is around $30\%$ \citep[see e.g.][for a review]{Winters2015}. Close companions can affect a star's rotation through tidal synchronization or disruption of the protostellar disk \citep[e.g.][]{Meibom2005,Morgan2012}. Unresolved multiples or background objects could also result in spurious period detections. Our tables note objects identified as multiples in the literature or by visual inspection, and any objects where the MEarth photometric aperture contained known background sources. Including both bright and faint companions, $230$ objects in our sample have a nearby, physically-associated companion and $449$ have a source in the background.
We additionally note objects identified as potentially over-luminous in our previous work. These were identified by \citet{Newton2015}, on the basis of their absolute magnitudes and spectroscopically-derived luminosities, or by \citet{Dittmann2015}, using their absolute magnitudes and colors. 

The analysis in this paper excludes the objects that have bright, unresolved companions (regardless of whether they are common proper motion or background objects, contamination flag $1$), or have been identified as over-luminous (flag $4$). This excludes $211$ of the $230$ objects with a nearby, physically-associated companion ($12\%$ of the full sample). The contaminants are distributed proportionally across the four possible period detection ratings. Objects with faint companions are not excluded. 

2MASS J11470543+7001588 (G 236-81) is one such multiple, and is the only object in which we clearly detect two unrelated periods. The periods are $3.49$ days and $5.37$ days.

\subsection{Defining a statistical sample}

For questions that seek to address how the rotators are different from the non-rotators, we need to know whether or not we could have detected rotation with a certain period and amplitude in a given star. The full Monte Carlo simulation necessary to adequately address period recovery is computationally intensive, so we instead use global properties of the light curves to define a ``statistical sample'' of stars for which we believe we could have detected long rotation periods. We find that a combination of the number of visits ($n_\mathrm{visits}$) to an object and the typical error ($\sigma$) of each visit is strongly predictive of whether or not we detected a rotation period, leading us to define the statistical sample as all stars with $n_\mathrm{visits}>1200$ and $\sigma < 0.005$ mag, where $\sigma$ is the median theoretical error divided by the square root of the number of exposures per visit. There are $223$ stars in the statistical sample.

\begin{figure}
\includegraphics[width=\linewidth]{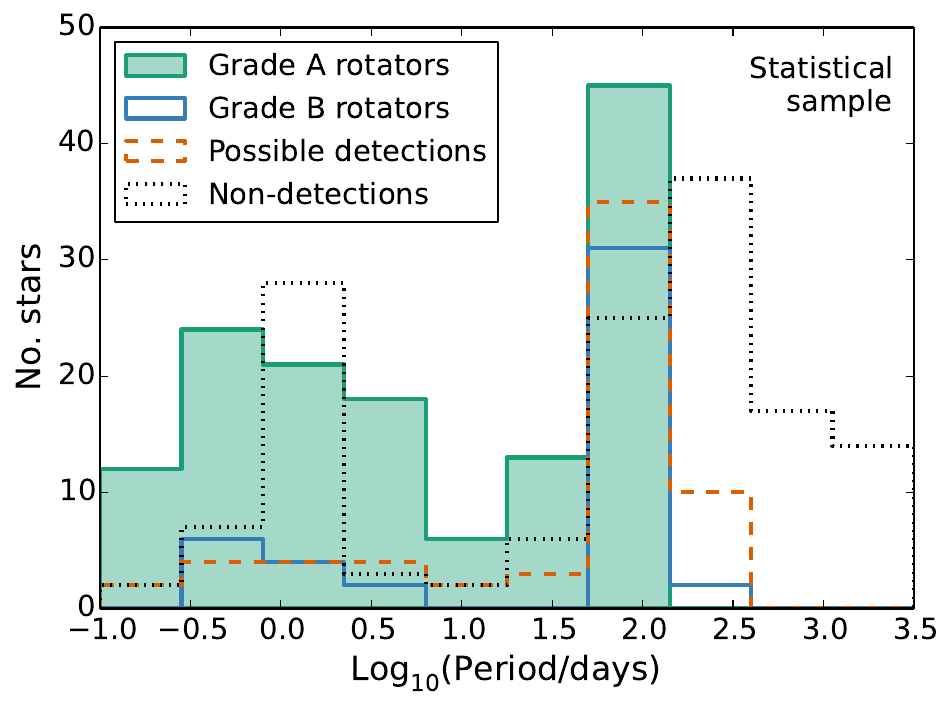}
\caption{Distribution of best-fitting periods for the grade A and grade B rotators, and uncertain detections in our statistical sample. The lack of grade A rotators at long periods is likely a result of incompleteness. We see a relative lack of stars with intermediate rotation periods, which we suggest is astrophysical in origin.
\label{Fig:distribution}}
\end{figure}

We show the distribution of periods in the statistical sample in Figure \ref{Fig:distribution}. The grade A rotators are biased towards shorter periods, which are easier to positively identify as being the result of stellar variability even within the statistical sample. The primary reason is that a short-period rotator undergoes more rotation cycles in a given amount of time than does a long-period rotator. The multiple rotation cycles seen for short periods helps to confirm low-amplitude signals in noisy data, and causes a greater fraction of stars to have enough data to satisfy our requirement of two cycles of modulation (see \S \ref{Sec:method_period}). The tendency for the grade B rotators and candidate rotators to have long periods is therefore due to the incompleteness of grade A rotators at long periods. Non-detections favor non-astrophysical periods near $1$ day or that are very long ($1000$ days).

We see a relative lack of stars with intermediate rotation periods around $30$ days, which we suggest is astrophysical in origin. In the statistical sample, the distribution of best-fitting periods for all stars (including possible detections and stars with no detection) does not indicate a large population of intermediate rotators. This lends support to the idea that our by-eye classification is not lacking sensitivity to intermediate-period rotators.

\section{Comparison to previous period measurements}\label{Sec:comparison}

A few dozen of our stars have been the targets of other surveys. In this section, we take a closer look at these objects. We first compare our work to that of other ground-based photometric surveys (\S\ref{Sec:ground}). In \S\ref{Sec:kepler}, we look at the few MEarth objects with photometry from the \kepler\ space telescope. The rotational broadening of spectral features provides another means to determine stellar rotation, and we present a comparison to those studies in \S\ref{Sec:vsini}.

\subsection{Comparison to ground-based photometry}\label{Sec:ground}

\begin{figure}
\begin{center}
\includegraphics[width=0.8\linewidth]{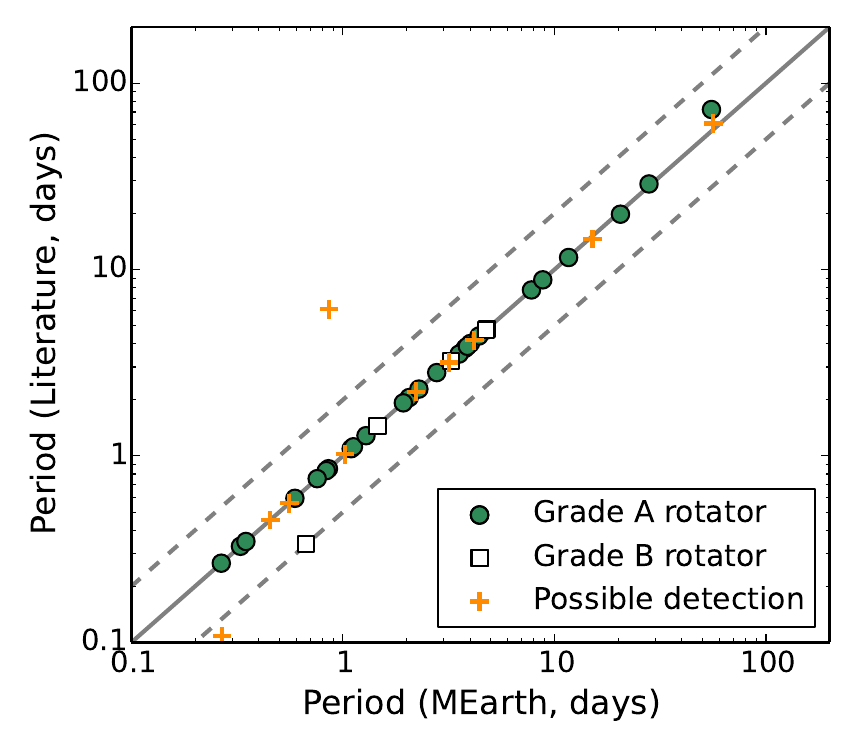}
\end{center}
\caption{Comparison between photometric period measurements from this work (horizontal axis) and literature sources (vertical axis). We indicate the MEarth period rating assigned for each star with different symbols: grade A rotators (filled circles), grade B rotators (open squares), and possible detections (plus symbols). The solid line indicates exact agreement, while the dashed lines indicate periods twice or half of what we measure. The strong outlier with a literature period at $6.11$ days likely results from the daily sampling alias.
\label{Fig:ground}}
\end{figure}

We compare our grade A and B rotators to those with photometric periods from the literature. We additionally show objects with ``possible'' ratings. Including known multiples, we have overlap with \citet[][1 star in common]{Alekseev1998}, \citet[][5 stars]{Norton2007}, \citet[][3 stars]{Hartman2010}, \citet[][2 stars]{Shkolnik2010}, and \citet[][25 stars]{Hartman2011}. 

\citet{Hartman2011}, our primary source for literature measurements, used data from HATNet that spanned time baselines of $45$ days to $2.5$ years. 
They searched for rotation periods between $0.1$ and $100$ days amongst all field K and M dwarfs using analysis of variance (AoV), which tries to find the period that minimizes the scatter in the phased light curve. They decorrelate against external parameters (``EPD'' light curves), then against templates built from other objects in the field (``TFA'' light curves), and report a quality flag for each detection. 
We exclude quality flags of $2$, and by default use the TFA-based detections. We adopt the EPD-based detections if they have a better quality flag. We noticed that for 2MASS J17195298+2630026 (Gl~669~B), the two algorithms resulted in different periods. The TFA analysis gives $P=1.45$ days, which agrees with the period we detect. The EPD analysis gives $P=20$ days, the same period as both we and \citet{Hartman2011} determine for the common proper motion companion 2MASS J17195422+2630030 (Gl~669~A). Although the EPD period had the higher quality flag, we adopt the TFA period for this object.

We find excellent agreement between the periods measured from these surveys and the periods we measure from MEarth (Figure \ref{Fig:ground}). Three objects are discrepant. For grade A rotator 2MASS J13505181+3644168 (LHS~6261), our period is $55.7$ days while that from \citet{Hartman2011} is $72.2$ days. We also detect a second strong frequency in this object with a period of $93$ days. Our frequencies evenly bracket the Hartman et al.\ detection, so we suspect our periodogram peak has been split as a result of the window function of the MEarth data.
For our ``possible''-rated object 2MASS J14545496+4108480 (LP~222-15), we find a period of $0.858$ days, while \citet{Hartman2011} measure a period of $6.11$ days. There is a peak in our periodogram at $6.2$ days, and $0.858$ is close to the one-day sampling alias of this signal. 
For our ``possible''-rated object 2MASS J13314666+2916368 (DG~CVn), we measure $P=0.268$ days, while \citet{Robb1999} measure $0.108$ days. Our candidate signal is affected both by the sparse data set and the baseline magnitude changes.

\begin{deluxetable*}{l r r r l}
\tablecaption{\label{Tab:ground}Objects with ground-based photometric periods from the literature}
\tablecolumns{7}
\tablehead{ 
	\colhead{2MASS ID} & 
	\colhead{Grade\tablenotemark{a}} & 
	\colhead{MEarth $P$} &
	\colhead{Lit $P$} &
	\colhead{Ref.\tablenotemark{b}}\\
	\colhead{} &
	\colhead{} &
	\colhead{(days)} &
	\colhead{(days)} &
	\colhead{}
	}
\startdata
\sidehead{Rotators}
\hline \\ 
J00285391+5022330 & A &  1.093 & 1.09332 & H11 \\ 
J02024428+1334335 & A &  4.003 & 4.01 & S10 \\ 
J03223165+2858291 & A &  1.929 & 1.92673 & H11 \\ 
J03364083+0329194 & A &  0.328 & 0.32766 & K12 \\ 
J03425325+2326495 & A &  0.834 & 0.834379 & H10 \\ 
J07382951+2400088 & A &  3.875 & 3.87463 & H11 \\ 
J07444018+0333089 & A &  2.775 & 2.8 & A98 \\ 
J08065532+4217333 & A &  8.804 & 8.80699 & H11 \\ 
J09214911+4330284 & A & 27.984 & 28.7811 & H11 \\ 
J09441580+4725546 & A &  4.395 & 4.40041 & H11 \\ 
J09591880+4350256 & A &  0.755 & 0.7554 & H11 \\ 
J10512059+3607255 & A &  3.782 & 3.77885 & H11 \\ 
J11031000+3639085 & A &  2.056 & 2.05692 & H11 \\ 
J11115176+3332111 & A &  7.785 & 7.77026 & H11 \\ 
J11474074+0015201 & A & 11.662 & 11.603 & K12 \\ 
J13505181+3644168 & A & 55.239 & 72.1768 & H11 \\ 
J15553178+3512028 & A &  3.542 & 3.52093 & H11 \\ 
J17195422+2630030 & A & 20.511 & 19.8077 & N07 \\ 
J17335314+1655129 & A &  0.266 & 0.2659 & N07 \\ 
J18130657+2601519 & A &  2.285 & 2.2838 & N07 \\ 
J19510930+4628598 & A &  0.593 & 0.592578 & H11 \\ 
J20103444+0632140 & A &  1.121 & 1.12 & S10 \\ 
J21322198+2433419 & A &  4.747 & 4.7358 & N07 \\ 
J22232904+3227334 & A &  0.854 & 0.854 & M11 \\ 
J23025250+4338157 & A &  0.348 & 0.347704 & H11 \\ 
J23050871+4517318 & A &  1.285 & 1.28447 & H11 \\ 
J00161455+1951385 & B &  4.798 & 4.7901 & N07 \\ 
J03284958+2629122 & B &  3.235 & 3.23062 & H10 \\ 
J04381255+2813001 & B &  0.670 & 0.335985 & H11 \\ 
J17195298+2630026 & B &  1.457 & 1.454184 & H11\tablenotemark{c} \\ 
J23545147+3831363 & B &  4.755 & 4.757 & K13 \\ 
\sidehead{Candidates}
\hline \\ 
J02253841+3732339 & U & 15.135 & 14.6016 & H11 \\ 
J03264495+1914402 & U &  0.454 & 0.454016 & H10 \\ 
J10235185+4353332 & U & 56.311 & 60.7517 & H11 \\ 
J10382981+4831449 & U &  3.178 & 3.17243 & H11 \\ 
J13314666+2916368 & U &  0.268 & 0.10835 & R99 \\ 
J13374043+4807542 & U &  0.558 & 0.55754 & H11 \\ 
J14545496+4108480 & U &  0.858 & 6.11491 & H11 \\ 
J14592508+3618321 & U &  4.173 & 4.16904 & H11 \\ 
J15040626+4858538 & U &  1.022 & 1.02136 & H11 \\ 
J15192126+3403431 & U &  2.211 & 2.21031 & H11
\enddata
\tablenotetext{a}{MEarth period rating, see description in text.}
\tablenotetext{b}{Reference for literature photometric period.~
A98 = \cite{Alekseev1998};
R99 = \citet{Robb1999};
N07 = \cite{Norton2007};
H10 = \cite{Hartman2010};
S10 = \cite{Shkolnik2010};
H11 = \cite{Hartman2011};
M11 = \citet{Messina2011};
K12 = \citet{Kiraga2012};
K13 = \citet{Kiraga2013}
}
\tablenotetext{c}{Period from TFA light curve, see text for discussion.}
\end{deluxetable*}

\subsection{Comparison to \kepler}\label{Sec:kepler}

\begin{figure}
\includegraphics[width=0.9\linewidth]{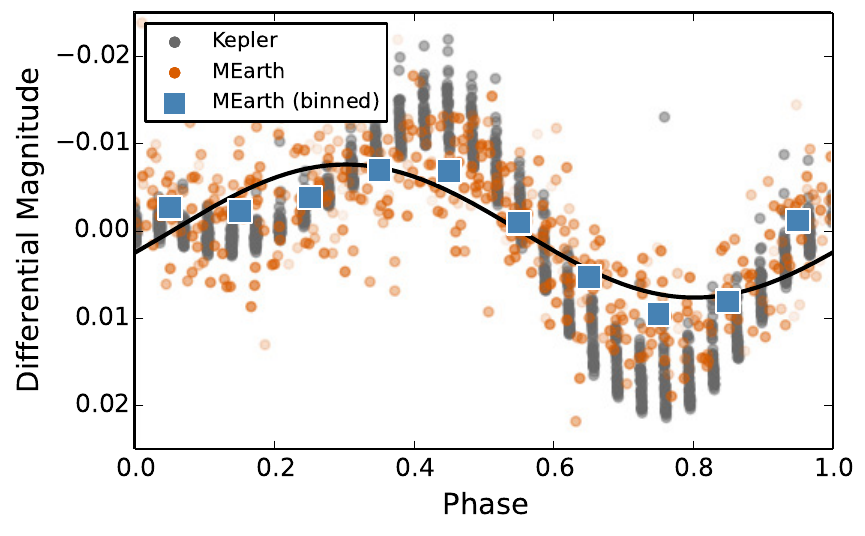}
\includegraphics[width=0.9\linewidth]{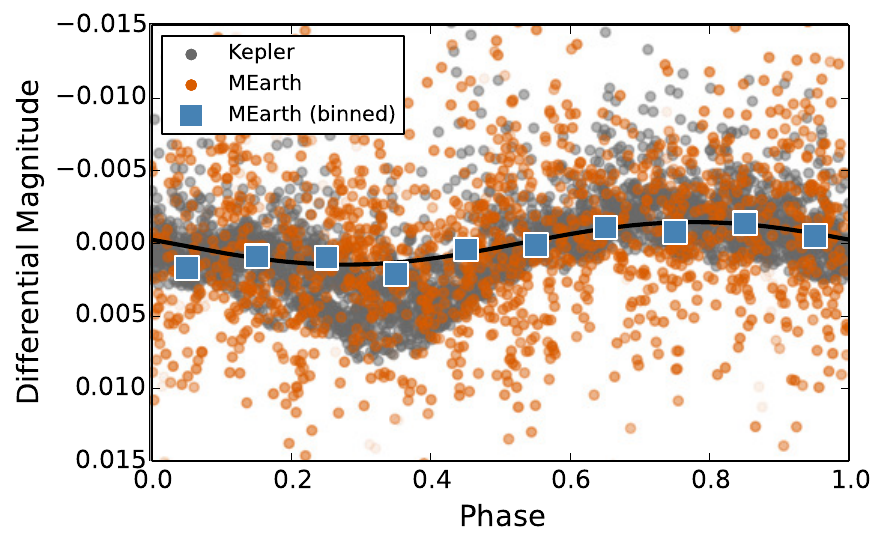}
\caption{Phased light curves for objects in common between the MEarth and \kepler\ samples, for which a rotation period was detectable by eye in \kepler. For KIC 9726699 (top), the rotation period is the best-fitting one found by our analysis. For KIC 9201463 (bottom), the period was identified first in the \kepler\ light curve, after which we found the closest-fitting peak in the periodogram of the MEarth data. Grey points show a subset of the \kepler\ data, MEarth data from the same period of time are shown in orange (with increasing transparency indicating data with larger errors), and the binned MEarth data in blue. The black curve is the sinusoid that best fits the MEarth data.
\label{Fig:kepler}}
\end{figure}

The \kepler\ space telescope gathered multi-year photometry on approximately $150000$ stars, including several thousand M dwarfs, most of which are early Ms \citep{Borucki2010, Koch2010}. We downloaded \kepler\ light curves for the objects with simultaneous data from MEarth from the  Mikulski Archive for Space Telescopes (MAST). We use the PDCSAP\_FLUX data; this reduction was intended to remove instrumental systematics while retaining astrophysical variability \citep{Stumpe2012,Smith2012}. Ten targets in our sample also have data from \kepler. We examine the two that have periodic modulations in the \kepler\ data that are detectable by eye (Figure \ref{Fig:kepler}). We have not detected periods in the remaining objects, though we note that for KIC 6117602, we have a candidate detection of $80$ days which is at odds with the 0.67 day period reported by \citet{Rappaport2014}.

For KIC 9726699 (GJ 1243, Figure \ref{Fig:kepler}, top panel), we detect a period at 0.59 days, which we assigned grade A. The period, including the indications of asymmetry, is easily identifiable in \kepler\ photometry.
For KIC 9201463 (Figure \ref{Fig:kepler}, bottom panel), for which the \kepler\ light curve has a clear $5.5$ day signal, we did not detect a rotation period in the MEarth data alone. However, the MEarth data has power at this frequency, and the modulation matches the phase of \kepler\ signal. The MEarth bandpass is redder than that of \kepler, so we expect the amplitude to be lower in our data. The relatively small amplitude (0.5\% in the \kepler\ bandpass), somewhat non-sinusoidal modulation, and the frequent flaring are also likely to contribute to our inability to detect the signal independently in MEarth.

\subsection{Comparison with \vsini\ measurements}\label{Sec:vsini}

The rate at which a star spins can also be inferred by measuring the broadening of spectral features due to the rotational velocity ($v$) of its photosphere. Due to the unknown inclination $i$ only \vsini\ can usually be determined. We searched the literature for previous measurements of \vsini\ for M dwarfs in the \citet{Nutzman2008} sample. The papers in which we looked for matches are listed in Table \ref{Tab:vsini-compilation}. We first compare \vsini\ measurements directly to other \vsini\ measurements from the literature (\S\ref{Sec:internal}), and use lessons from this analysis in our comparison of \vsini\ and photometric rotation period (\S\ref{Sec:external}).

\begin{deluxetable*}{lll}
\tablecaption{\label{Tab:vsini-compilation}Sources for \vsini\ compilation}
\tablecolumns{3}
\tablehead{
	\colhead{Abbreviation} &
	\colhead{Reference} &
	\colhead{Resolving power} \\
	\colhead{} &
	\colhead{} & 
	\colhead{($R/1000$)}
	}
\startdata
V83 & \citet{Vogt1983} &             115  \\
SH86 & \citet{Stauffer1986} &      20   \\
S87 & \citet{Stauffer1987} &         20  \\
MC92 & \citet{Marcy1992} &             40   \\
T92 & \citet{Tokovinin1992} &         18   \\
JV96 & \citet{Johns-Krull1996} &    120   \\
S97 & \citet{Stauffer1997} &           44    \\
D98 & \citet{Delfosse1998} &          42   \\
TR98 & \citet{Tinney1998} &               19   \\
B00 & \citet{Basri2000} &              31    \\
S01 & \citet{Schweitzer2001} &         34/45   \\
G02 & \citet{Gizis2002} &               19  \\
R02 & \citet{Reid2002} &               33   \\
MB03 & \citet{Mohanty2003} &             31  \\
B04 & \citet{Bailer-Jones2004} &         39   \\
FS04 & \citet{Fuhrmeister2004} &       45   \\
J05 & \citet{Jones2005} &              42   \\
Z05 & \citet{Zickgraf2005} &            20/22/34    \\
T06 & \citet{Torres2006} &              50    \\
Z06 & \citet{ZapateroOsorio2006} &      20  \\
R07 & \citet{Reiners2007} &             200   \\
RB07 & \citet{Reiners2007} &             31   \\
H07 & \citet{Houdebine2008} &          45     \\
RB08 & \citet{Reiners2008} &            31/33   \\
J09 & \citet{Jenkins2009} &              37    \\
WB09 & \citet{West2009} &                 31    \\
B10 & \citet{Blake2010} &                25   \\
H10 & \citet{Houdebine2010} &           40/42/75   \\
RB10 & \citet{Reiners2010} &            31/32  \\
B10 & \citet{Browning2010} &           60   \\
R12 & \citet{Reiners2012a} &            40/48  \\
Bai12 & \citet{Bailey2012} &              30   \\
Bar12 & \citet{Barnes2012} &              35 \\
D12 & \citet{Deshpande2012} &          20   \\
H12 & \citet{Houdebine2012} &          75/115  \\
K12 & \citet{Konopacky2012} &              30   \\
T12 & \citet{Tanner2012} &               24   \\
D13 & \citet{Deshpande2013} &            22.5    \\
M13 & \citet{Mamajek2013} & 		100 \\
B14 & \citet{Barnes2014} &                54   \\
M14 &  \citet{Malo2014} & 	50/52/68/80  \\ 
D15 & \citet{Davison2015} &                57   \\
HM15 & \citet{Houdebine2015} &          75/115 
\enddata
\end{deluxetable*}

\subsubsection{Comparison between literature \vsini\ measurements}\label{Sec:internal}

Several of the surveys with \vsini\ measurements for our targets used spectrographs with resolving powers ($R\equiv\lambda/\Delta\lambda$) of $20000$ to $40000$. At these resolutions, the rotational broadening of all but the most rapid rotators falls below the resolution of the spectrograph, and disentangling non-astrophysical sources of broadening from rotational broadening is challenging. 

\begin{figure}
\includegraphics[width=\linewidth]{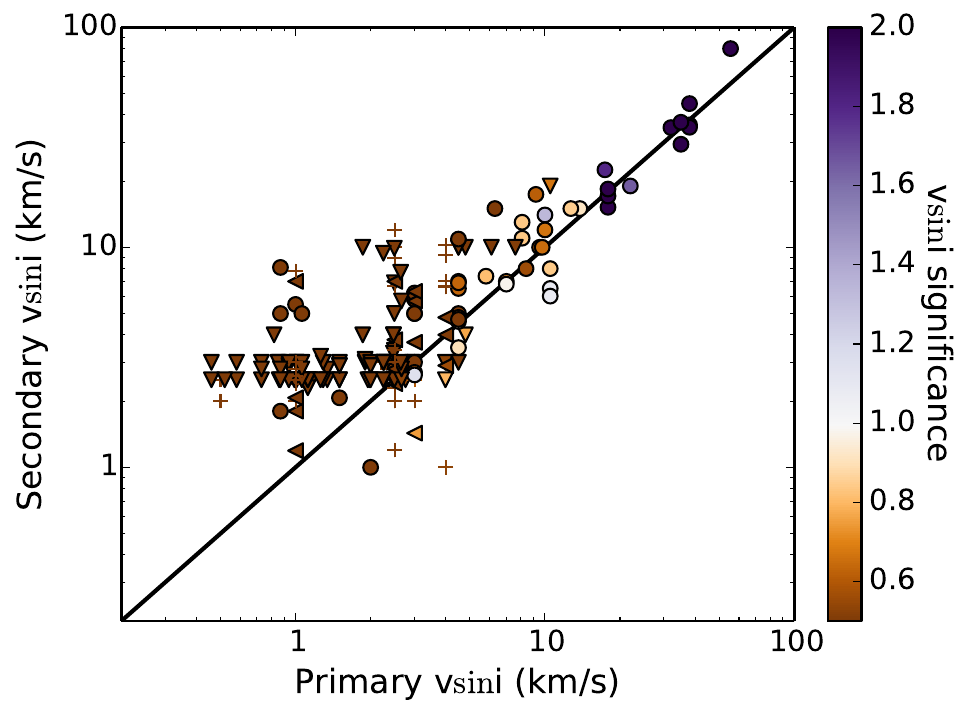}
\caption{Comparison of \vsini\ measurements from the literature for M dwarfs in our database. The horizontal axis shows the \vsini\ measured by the highest-spectral resolution survey (requiring $R>40000$, called the ``primary'' survey), while the vertical axis shows the \vsini\ measured from other (``secondary'') surveys. The color indicates our estimate of the significance of the detection; smaller values (brown) are less significant, while larger values (purple) are more so. The symbol shape indicates whether the reported value is a detection or an upper limit: Solid circles indicate detections reported by both the primary and secondary surveys. Triangles pointed downwards indicate that the secondary survey reported an upper limit, while triangles pointed towards the left indicate that the primary survey reported an upper limit. Plus symbols indicate that both the primary and secondary survey report upper limits.
\label{Fig:vsini-lit}}
\end{figure}

We directly compare \vsini\ for M dwarfs from different literature sources, regardless of whether the star was observed by MEarth, in Figure \ref{Fig:vsini-lit}. Since we do not know the true \vsini, we consider the value measured by the highest-spectral resolution survey (which we call the ``primary'' survey) and compare it to values measured by lower-spectral resolution (``secondary'') surveys. We require that the primary survey resolution be greater than $40000$. 

For $v\sin{i}>20$ km/s, the primary and secondary surveys do not deviate systematically, but there are significant discrepancies for smaller values of \vsini. We find that when inferring \vsini\ broadening that is below the spectral resolution, the secondary surveys tend to determine higher values for \vsini\ than the primary survey, and the magnitude of the discrepancy varies with the significance of the reported detection. Here we define the \vsini\ significance as the \vsini\ measured by the primary survey divided by the resolving power in km/s of the spectrograph used in the secondary survey. We find discrepancies in many \vsini\ detections with significances less than about $0.8$.  We therefore arrive at a similar conclusion to \citet{Reiners2012a}, who found evidence that some detections of \vsini\ are spurious.\footnote{ \citet{Houdebine2015} carried out a comparison of \vsini\ values and found their measurements and other surveys agreed well. However, their comparison did not include all of the lower-resolution surveys we considered.} Improved treatment of \vsini\ detection limits, as well as additional or repeated \vsini\ measurements at higher resolving power, would be beneficial.

\subsubsection{Comparison between \vsini\ and photometric measurements}\label{Sec:external}

\begin{figure}
\includegraphics[width=\linewidth]{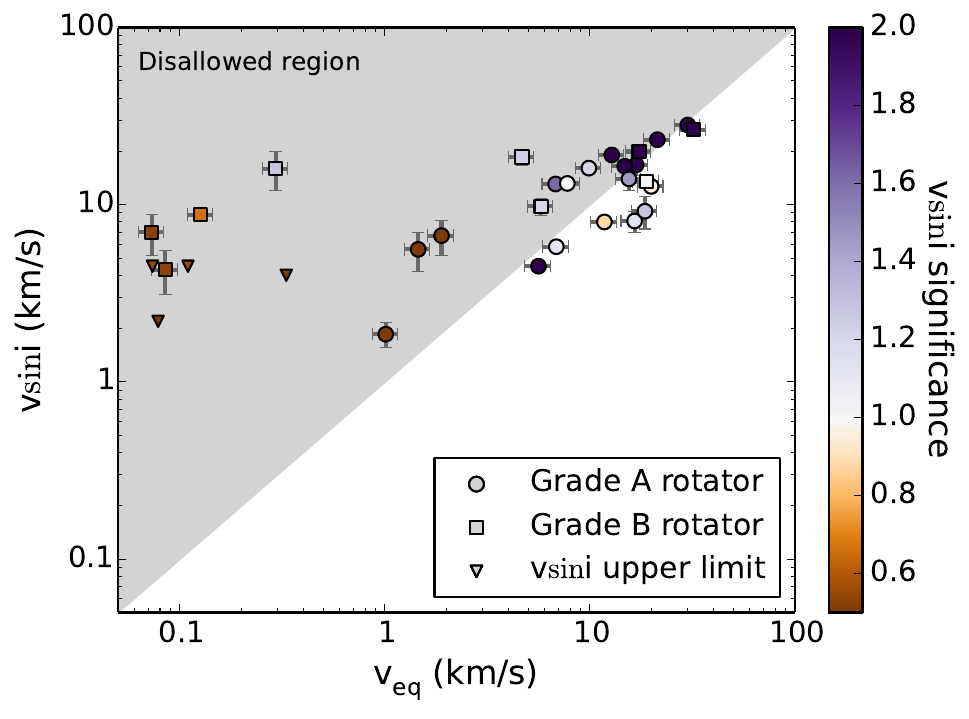}
\caption{Comparison of estimated equatorial rotation velocities (horizontal axis) and \vsini\ measurements from the literature. The color indicates our estimate of the significance of the detection, assuming the \vsini\ reported and the resolution of the spectrograph used; smaller values (brown) are less significant, while larger values (purple) are more so. Solid circles (for grade A rotators) and squares (for grade B rotations) indicate \vsini\ detections, while triangles indicate an upper limit. Errors on \vsini\ are included where available, and we have estimated errors for $v_{eq}$. The gray shaded region indicates the region where $v\sin{i}>v_\mathrm{eq}$; no detections should fall in this region.
\label{Fig:vsini}}
\end{figure}

We compare the equatorial rotational velocities ($v_\mathrm{eq}$) we infer from the photometric rotation period to the measured \vsini\ in Figure \ref{Fig:vsini} and Table \ref{Tab:vsini}. This comparison includes our grade A and B rotators. $v_\mathrm{eq}$ is calculated from the estimated stellar radius ($R$) and the rotation period ($P$) by:
\begin{align}
v_\mathrm{eq} = \frac{2\pi R}{P}
\end{align}
We do not calculate errors on the rotation period, so for this comparison we assume a $10\%$ error on period \citep[]{Irwin2011} and a $10\%$ error on stellar radius \citep[]{Delfosse2000, Boyajian2012}. If the photometric rotation period, stellar radius, and the \vsini\ are correct, $v_\mathrm{eq}>v\sin{i}$. Significance is defined as before: the reported \vsini\ measurement divided by the resolving power of the spectrograph used.

\begin{deluxetable*}{l c r r r r l}
\tablecaption{\label{Tab:vsini}Rotators with \vsini\ measurements}
\tablecolumns{7}
\tablehead{ 
	\colhead{2MASS ID} & 
	\colhead{Grade} & 
	\colhead{MEarth $P$\tablenotemark{a}} &
	\colhead{$v_{eq}$\tablenotemark{b}}&
	\colhead{\vsini} &
	\colhead{Resolving power} &
	\colhead{Ref.\tablenotemark{d}} \\
	\colhead{} &
	\colhead{} &
	\colhead{(days)} &
	\colhead{(km/s)} &
	\colhead{(km/s)} &
	\colhead{($R/1000$)} & 
	\colhead{}
	}
\startdata
J02170993+3526330 & A   &   0.3 & $30.0\pm4.2$ & $28.2\pm0.7$ &  37 & J09 \\ 
J02204625+0258375 & A   &   0.5 & $21.3\pm3.0$ & $23.3\pm 0.7$ &  37 & J09 \\ 
J03205965+1854233 & A   &   0.6 & $11.8\pm1.7$ & $8.0\pm\cdots$ &  33 & R02 \\ 
J03425325+2326495 & A   &   0.8 & $19.9\pm2.8$ & $12.7\pm0.5$ &  22.5 & D13 \\ 
J06000351+0242236 & A   &   1.8 &  $6.9\pm1.0$ & $5.8\pm0.3$ &  57 & D15 \\ 
J07444018+0333089 & A   &   2.8 &  $5.6\pm0.8$ & $4.5\pm\cdots$ & 200 & R07 \\ 
J08294949+2646348 & A   &   0.5 & $16.5\pm2.3$ & $8.1\pm1.1$ &  42 & D98 \\ 
J08505062+5253462 & A   &   1.8 &  $6.8\pm1.0$ & $13.1\pm0.7$ &  37 & J09 \\ 
J09301445+2630250 & A   &  10.7 &  $1.9\pm0.3$ & $6.7\pm1.5$ &  22.5 & D13 \\ 
J09535523+2056460 & A   &   0.6 & $14.8\pm2.1$ & $16.5\pm0.4$ &  37 & J09 \\ 
J10163470+2751497 & A   &  22.0 &  $0.33\pm0.05$ & $<4.0$ &  33 & R02 \\ 
J10521423+0555098 & A   &   0.7 & $12.8\pm1.8$ & $19.1\pm0.2$ &  37 & J09 \\ 
J11474074+0015201 & A   &  11.7 &  $1.5\pm0.2$ & $5.6\pm1.4$ &  22.5 & D13 \\ 
J12185939+1107338 & A   &   0.5 & $18.6\pm2.6$ & $9.2\pm1.9$ &  42 & D98 \\ 
J13003350+0541081 & A   &   0.6 & $16.8\pm2.4$ & $16.8\pm2.1$ &  42 & D98 \\ 
J13564148+4342587 & A   &   0.5 & $15.5\pm2.2$ & $14.0\pm2.0$ &  31 & R10 \\ 
J17195422+2630030 & A   &  20.5 &  $1.0\pm0.1$ & $1.9\pm0.3$ &  75 & H12 \\ 
J18024624+3731048 & A   & 123.8 &  $0.07\pm0.01$ & $<4.5$ &  37 & J09 \\ 
J18452147+0711584 & A   &   0.8 &  $9.9\pm1.4$ & $16.1\pm0.1$ &  22.5 & D13 \\ 
J19173151+2833147 & A   &   1.1 &  $7.8\pm1.1$ & $13.2\pm0.5$ &  22.5 & D13 \\ 
J22245593+5200190 & A   &  81.8 &  $0.11\pm0.02$ & $<4.5$ &  37 & J09 \\ 
J03132299+0446293 & B   & 126.2 &  $0.08\pm0.01$ & $<2.2$ &  42 & D98 \\ 
J05011802+2237015 & B   &  70.7 &  $0.13\pm0.02$ & $8.8\pm0.3$ &  22.5 & D13 \\ 
J06022918+4951561 & B   & 104.6 &  $0.08\pm0.01$ & $4.3\pm1.2$ &  37 & J09 \\ 
J09002359+2150054 & B   &   0.4 & $17.4\pm2.5$ & $20.0\pm0.6$ &  37 & J09 \\ 
J11005043+1204108 & B  &   0.3 & $31.9\pm4.5$ & $26.5\pm0.8$ &  22.5 & D13 \\ 
J12265737+2700536 & B   &   0.7 & $18.8\pm2.7$ & $13.5\pm0.6$ &  22.5 & D13 \\ 
J16370146+3535456 & B   & 100.4 &  $0.07\pm0.01$ & $7.0\pm1.8$ &  22.5 & D13 \\ 
J22081254+1036420 & B   &   2.4 &  $4.7\pm0.7$ & $18.6\pm2.0$ &  20 & D12 \\ 
J23134727+2117294 & B   &  34.5 &  $0.29\pm0.04$ & $16.0\pm4.0$ &  24 & T12  \\
J23354132+0611205 & B   &   1.7 &  $5.8\pm0.8$ & $9.8\pm1.1$ &  37 & J09
\enddata
\tablenotetext{a}{Rotation period determined from MEarth photometry in this work.}
\tablenotetext{b}{Equatorial rotation velocity and error calculated from the MEarth rotation period and the estimated stellar radius, assuming $10\%$ errors on both the radius and the period.}
\tablenotetext{c}{Resolving power of the spectrograph used in the \vsini\ study, divided by $1000$.}
\tablenotetext{d}{Reference code for \vsini\ measurement.  See Table \ref{Tab:vsini-compilation}.}
\end{deluxetable*}

We first note the stars with reported \vsini\ detections at low significance, for which the \vsini\ broadening is less than the resolving power of the spectrograph used (brown circles and squares in Figure \ref{Fig:vsini}). This includes the three objects for which we measured long photometric rotation periods but that have \vsini\ measurements indicating rapid rotation. Based on the analysis we presented previously in Figure \ref{Fig:vsini-lit}, these low-significance \vsini\ detections may be incorrect, and we suggest this is the cause of the disagreement with our results. 

We will now consider only \vsini\ detections at higher significance (white and purple points in Figure \ref{Fig:vsini}). Our photometric periods concur with the detection of rapid spin from rotational broadening. However, \vsini\ still exceeds $v_\mathrm{eq}$ in some cases, so one or more of the \vsini, our photometric period, or the radius estimate must be in error. In some cases, the highest peak in the periodogram may represent an alias or harmonic of the true period of the star. The most extreme example is 2MASS J23134727+2117294 (LP~462-11), with $v\sin{i}=16$ km/s and $P=34.5$ days ($v_\mathrm{eq} = 0.3$ km/s). Our light curve for this object is relatively sparse (which engendered it a grade B rating); and although the strongest signal in the periodogram is at $P=34.5$, a rotation period of close to one day also provides a reasonable fit to the data. Table \ref{Tab:update} considers stars with $v\sin{i}>v_\mathrm{eq}$ and provides the strongest signal at periods shorter than the best-fitting period.

\begin{deluxetable*}{l c r r r r r r r}
\tablecaption{\label{Tab:update}Alternate short-period signals for stars with $v\sin{i}>v_\mathrm{eq}$}
\tablecolumns{9}
\tablehead{ 
	\colhead{2MASS ID} & 
	\colhead{Grade} & 
	\colhead{MEarth $P$\tablenotemark{a}} &
	\colhead{$v_{eq}$\tablenotemark{a}} &
	\colhead{F-stat} &
	\colhead{Alt. $P$\tablenotemark{b}} &
	\colhead{Alt. $v_{eq}$} &
	\colhead{Alt. F-stat} &
	\colhead{\vsini}	\\
	\colhead{} &
	\colhead{} &
	\colhead{(days)} &
	\colhead{(km/s)} &
	\colhead{} &
	\colhead{(days)} &
	\colhead{(km/s)} &
	\colhead{} &		
	\colhead{(km/s)}
	}
\startdata
J02204625+0258375 & A        &   0.5 & 21.4 & 270 & 0.33 & 32.4 & 180 & 28.2  \\
J08505062+5253462 & A        &   1.8 &  6.8 & 250 & 0.64 & 19.1 & 140 & 13.1 \\
J09301445+2630250 & A        &  10.7 &  1.9 & 160 & 0.91 & 22.3 & 100 & 6.7 \\
J09535523+2056460 & A        &   0.6 & 14.8 &  75 & 0.38 & 23.4 & 60 & 16.5 \\
J10521423+0555098 & A        &   0.7 & 12.8 & 360 & 0.41 & 21.9 & 250 & 19.1 \\
J11474074+0015201 & A        &  11.7 &  1.5 & 540 & 1.09 & 16.1 & 340 & 5.6 \\
J17195422+2630030 & A        &  20.5 &  1.0 & 740 & 1.05 & 19.5 & 540 & 1.9 \\
J18452147+0711584 & A        &   0.8 &  9.9 & 150 & 0.46 & 17.2 & 110 & 16.1 \\
J19173151+2833147 & A        &   1.1 &  7.8 & 80 & 0.53 & 16.2 & 50 & 13.2 \\
J22081254+1036420 & B        &   2.4 &  4.7 & 180 & 0.7 & 16.1 & 180 & 18.6  \\
J23134727+2117294 & B        &  34.5 &  0.3 & 80 & 1.03 & 10.0 & 50 & 16.0 \\
J23354132+0611205 & B        &   1.7 &  5.8 & 40 & 0.62 & 15.9 & 30 & 9.8
\enddata
\tablenotetext{a}{Rotation period and $v_{eq}$ determined from MEarth photometry, reproduced from Table \ref{Tab:vsini} for clarity.}
\tablenotetext{b}{Alternative periods with power in the MEarth data. These periods are less significant than the adopted period.}
\end{deluxetable*}

\section{Spot characteristics}

We consider different aspects of starspots in this section. First, we investigate the relationship between semi-amplitude and rotation period (\S\ref{Sec:per-amp}), then consider spot patterns and the stability of the photometric modulations (\S\ref{Sec:fastslow}). In \S\ref{Sec:recovery}, we compare the fraction of stars that we detect to be rotating to the recovery rates of photometric and \vsini\ surveys.

\subsection{Amplitude of variability}\label{Sec:per-amp}

The amplitude of the photometric modulation is derived from the combined effect of the contrast between the spotted and unspotted stellar photosphere and the longitudinal inhomogeneity in the distribution of spots. Starspots, in turn, are surface manifestations of a star's magnetic field. Because rotation, magnetic fields, and starspots are closely related, we might therefore expect a correlation between rotation period and amplitude. 

\citet{Hartman2011} found that the rotation periods and amplitudes of K and early- to mid-M dwarfs are uncorrelated for periods less than $30$ days (see Figure 16 in their work), and that amplitude decreases with increasing rotation period for $P>30$ days. For later M dwarfs, \citet{Hartman2011} similarly found no correlation between amplitude and period for periods of up to $30$ days, but their sample contained few objects at longer periods. An anti-correlation was also seen by \citet{McQuillan2014} for M dwarfs in the \kepler\ sample; this sample is dominated by early M dwarfs and considered all stars with $\teff<4000$ K together. \citet{McQuillan2014} also identified a population of rotators with periods $<15$ days and high variability at a range of effective temperatures; examining Figure 14 in their work, the amplitude and period for this population appears uncorrelated. They postulate that these objects are binaries. 

We use semi-amplitudes in this analysis, where the semi-amplitude is defined as the coefficient of the best-fitting sinusoid ($a$ in Eq. \ref{Eq:sine}). Data on a single star may include several light curves whose semi-amplitudes are fit independently; the values we use are those from the light curve with the most observations. We derive errors using the covariance matrix from our least-squares fit, which does not include uncertainty in the period (see \S\ref{Sec:method}). The median amplitude error is $0.002$ magnitudes, and is independent of rotation period and the rating we have assigned for the period. Note that if the light curve is evolving or shows non-sinusoidal behavior, the semi-amplitude is suppressed relative to the peak-to-peak amplitude. 

\begin{figure}
\begin{center}
\includegraphics[width=\linewidth]{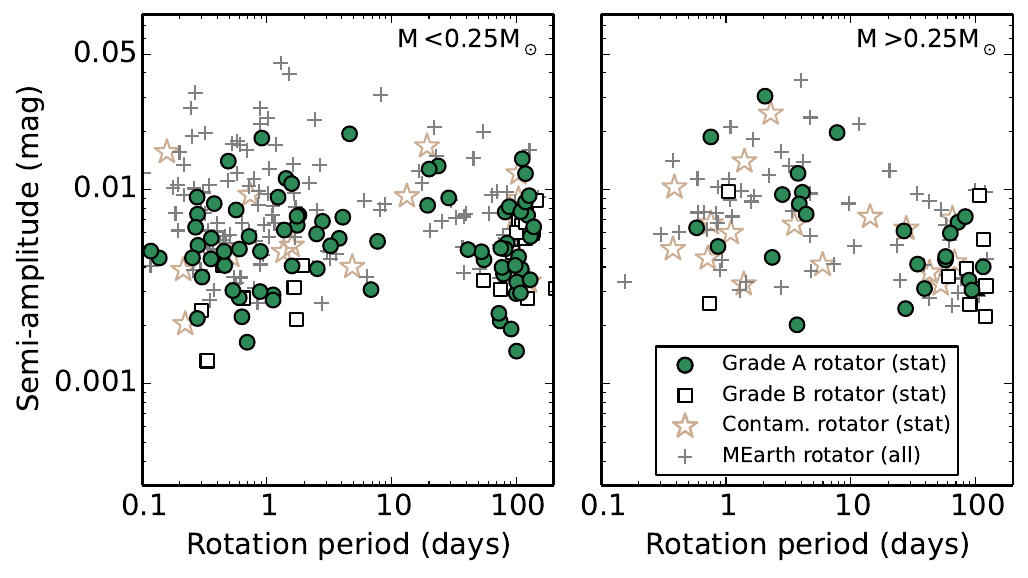}
\includegraphics[width=\linewidth]{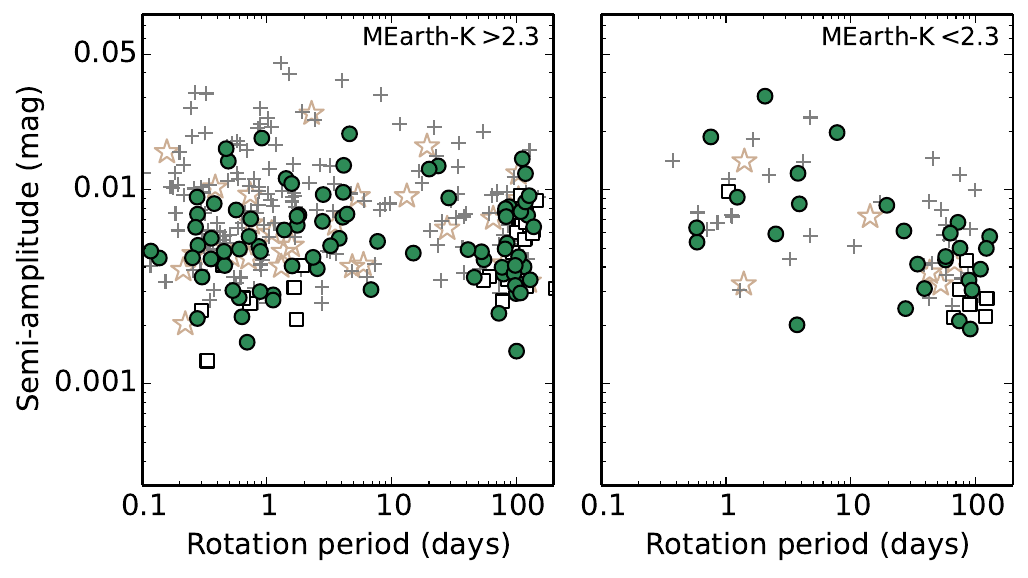}
\caption{Semi-amplitude of variability (defined as the coefficient of the sinusoid in our best fit) versus photometric rotation period, for $M<0.25\ \msun$ (left panel) and $M>0.25\ \msun$ (right panel). The median error on semi-amplitude is $0.002$ magnitudes. We caution that due to our use of absolute magnitudes to estimate stellar mass, a greater proportion of objects assigned to the higher mass bin are likely to be unresolved multiples. For objects in our statistical sample, we plot the grade A rotators (filled circles) and B rotators (open squares), and rotators with unresolved companions or that appear over-luminous (open stars). We also show, for reference, all rotators not in the statistical sample (plus signs). Our sensitivity to high-amplitude, short-period rotators in sparse data sets can be seen in the over-abundance of these objects in our full sample.
\label{Fig:per-amp}}
\end{center}
\end{figure}

In Figure \ref{Fig:per-amp}, we plot the semi-amplitude of variability versus rotation period for higher-mass ($0.25 < M \lesssim 0.5\ \msun$) and lower-mass ($0.08\lesssim M <0.25\ \msun$) M dwarfs in MEarth. Our mass limits approximate the $V-K$ color limits used by \citet{Hartman2011}. We show the divisions based on both stellar mass (estimated from absolute $K$ magnitudes) and color (using $MEarth-K$). We use our statistical sample in this analysis to avoid bias because high-amplitude, rapid rotators can be detected in many light curves where we are not sensitive to lower-amplitude or longer-period variables. 

We find a negative correlation between period and semi-amplitude for the higher-mass M dwarfs, but no correlation for mid M dwarfs, consistent with previous results. We use a Spearman rank correlation analysis to test the statistical significance of these results, and calculate the p-value for a two-sided hypothesis test with the null hypothesis that the data are uncorrelated, using the SciPy {\tt stats} package. Smaller p-values indicate higher confidence that the correlation is not due to chance. The correlation coefficient including both the grade A and B rotators is $-0.43\pm0.07$ for $M>0.25\ \msun$ ($p=0.01$) and $-0.01\pm0.03$ for $M<0.25\ \msun$ ($p=0.5$). Values reported are the median and $68\%$ confidence limits from a Monte Carlo simulation where we resampled with perturbation as suggested by \citet{Curran2014}. The lack of correlation for the lower mass stars also persists if we consider narrower ranges in mass. 

\subsection{Spot patterns and stability}\label{Sec:fastslow}

In keeping with our finding of the lack of a correlation between rotation and semi-amplitude, we find that most of our detected rotators show phase-folded light curves with qualitatively similar morphologies. 
At the precision of our data, they are usually sinusoidal in appearance. This could imply that the photometric modulations are the result of many spots acting in concert, or of a long-lived polar or high-latitude spots viewed at high inclination. Considering the former scenario, \citet{Jackson2013} demonstrated that photometric modulations of the amplitude we see can be produced by a large number of randomly distributed spots. The latter scenario is reflected in the prevalence of poloidal, axisymmetric large-scale fields recovered by Zeeman Doppler Imaging for fully-convective stars \citep[e.g.][]{Morin2008}, and in spot models from time series photometry or spectroscopy \citep[e.g][]{Davenport2015, Barnes2015}. These patterns tend to be stable over multiple rotation cycles, and in some cases over more than a year.

\begin{figure*}
\begin{center}
\begin{tabular}{ccc}
\includegraphics[width=\linewidth]{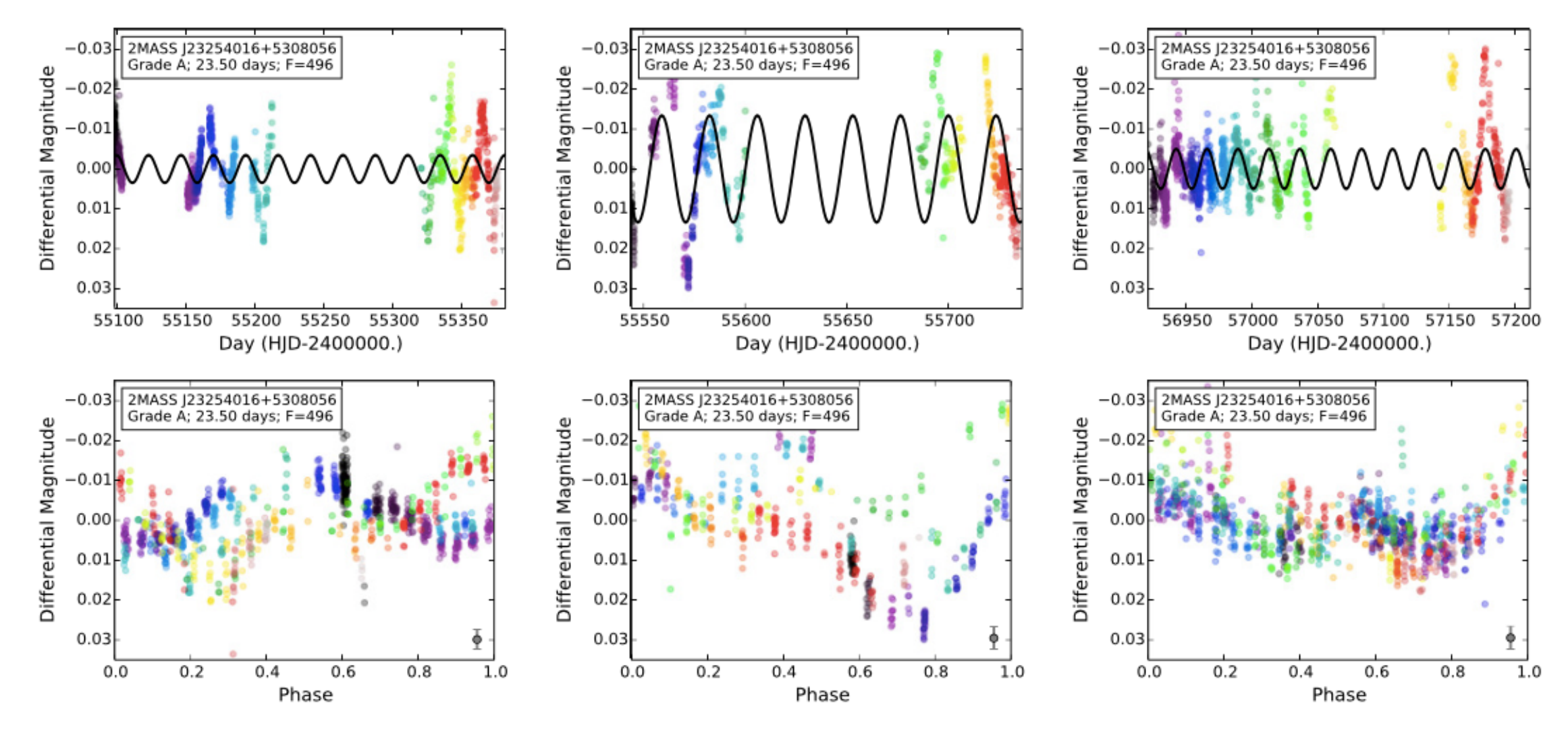}
\end{tabular}
\caption{Light curves of 2MASS J23254016+5308056 (LHS~543a) from 2008-2010 (left), 2010-2011 (center), and 2011-2015 (right). The top panels show the brightness as a function of time, with the best-fitting sinusoidal model over-plotted; the bottom panels show the light curves phase-folded to the best period. This object has the strongest and most rapid spot evolution of the stars in which we detect rotation periods. As in Figure \ref{Fig:examples}, the color of the data points indicates the observation epoch. The median error is represented in the bottom corner. \label{Fig:lspm1628}}
\end{center}
\end{figure*}

Aided by our visual inspections of the data, we are able to detect objects with evolving spot patterns. We highlight 2MASS J23254016+5308056 (LHS~543a) as the star demonstrating the strongest spot evolution in our sample. Light curves for this star, which we classify as a grade A rotator with a rotation period of $23.5$ days, are shown in Figure \ref{Fig:lspm1628}. The patterns seem stable for about two rotation cycles, and show evolution over roughly $200$ days.
We stress, however, that we expect our period detection method to be less effective for stars on which the spot patterns evolve on timescales comparable to the stellar rotation period.

Zeeman Doppler imaging measurements of late M dwarfs indicate that the magnetic field topologies of these stars can be very different for stars with similar properties. \citet{Donati2008} found that some late M dwarfs had axisymmetric, mostly dipolar fields (similar to earlier M dwarfs), while some are weaker, with more energy at small scales. We do not see any obvious dichotomy amongst the patterns of variability, but it is possible that one of the magnetic field topologies is more effective at producing spot contrast than the other.

\subsection{Recovery fractions}\label{Sec:recovery}

Previous photometric surveys have found a high fraction of fully-convective stars to be photometrically variable. \citet{McQuillan2014} find that approximately $80\%$ of the latest M dwarfs in \kepler\ have periods detected from their autocorrelation analysis, noting that their recovery of periods for these stars is not limited by the amplitude of variability. The recovery fraction of ground-based surveys is usually lower due to the cadence and precision of the observations. \citet{Hartman2011}, correcting for incompleteness using signal injection, estimate that $50\%$ of the stars with $M\lesssim0.2\msun$ are variable at semi-amplitudes $\gtrsim0.005$ mag in their bandpass (Cousins $I_C$ and $R_C$).

Our recovery rate of grade A and B rotators in the statistical sample is $47\pm5\%$, with no significant difference between the low- and high-mass populations. 
Considering $P<100$ days to match the period range studied by \citet{Hartman2011}, our recovery rate is $36\pm3\%$. The amplitude sensitivity of the two surveys is similar, but HATnet uses bluer photometric bandpasses where the contrast between the stellar photosphere and cooler spots is higher. We have also not modeled the incompleteness of our survey, though our use of the statistical sample mitigates the larger part of this effect for stars with $P<100$ days. 

Surveys of \vsini\ indicate a larger fraction of fully-convective stars are rotating rapidly than does our work. In a volume-limited survey, \cite{Delfosse1998} found that $50\%$ of field mid-M dwarfs (roughly M4V--M5V) are rotating rapidly enough to have detectable \vsini. \citet{Mohanty2003}, using new measurements and including those from \citet{Delfosse1998}, similarly found that half of their mid-M dwarf population (M4V-M5.5V) had detectable \vsini. In another survey, \citet{Browning2010} found that $30\%$ of M4.5V-M6V stars had detectable \vsini. For a $0.2\rsun$ star and a detection limit of $3$ km/s (typical for the two \vsini\ studies discussed here), this implies a period of less than $3.3$ days. We find that only $18\pm2\%$ of stars in our statistical are grade A or B rotators with $P<3.3$ days. 

The stellar samples selected by these surveys may not be comparable. For example, as the MEarth sample is proper motion selected, we are missing a larger fraction of stars with lower tangential velocities. These kinematically-cold stars are likely to be preferentially younger and therefore faster rotators (see \S\ref{Sec:unbiased}). We estimate that our sample represents $85$ to $90\%$ of the kinematically-unbiased sample (see \S\ref{Sec:unbiased}). If we add an additional $15\%$ of stars to our sample and assume that all are rotating at $P<3.3$ days, we can increase the fraction of rapid rotators to $30\%$. This would bring our results into agreement with those from \citet{Browning2010}, but still falls below the fractions reported by \citet{Delfosse1998} and \citet{Mohanty2003}.

Our photometric survey could have missed a population of short-period rotators: First, we know from KIC 9201463 that we are not able to detect all short-period rotators. Second, our method for period detection is not sensitive to stars whose spots evolve on timescales comparable the stellar rotation period. The roughly $30\%$ of rapid rotators we would need to have missed could be a population of rapidly-rotating stars with spot patterns that are not stable or that do not provide variability amplitudes high enough for us to detect. Aliasing of periods near $1$ day could also contribute (see \S\ref{Sec:vsini}).

\section{Kinematics and metallicities of the rotators}

To study kinematics, we require information on both the stars' positions and their motions through the Galaxy. As our targets were selected from a proper motion survey \citep{Lepine2005a}, all have measured proper motions. The majority of our targets also have parallaxes from MEarth astrometry measured by \citet{Dittmann2014}, though we use more precise measurements from the literature if available. We also gather radial velocities (RVs) from the literature, many of which come from \citet{Newton2014}, in which we used $R\approx2000$ near-infrared spectra to measure absolute RVs to $4$ km/s. This survey targeted many of the MEarth rotators that had been identified by the time of observation, but the availability of RVs still limits the fraction of stars for which we have kinematic information.

With all six phase space dimensions, we then calculate the $U$ (radial, positive is towards the Galactic center), $V$ (azimuthal), and $W$ (vertical) velocity components and their errors using an implementation of the method of \citet{Johnson1987}, updated to ICRS using the Galactic coordinate system defined in \citet[][vol. 1, part 1, sec. 1.5.3]{Perryman1997}. These velocities are measured relative to the Solar System barycenter. When we consider velocities relative to the local standard of rest, we will denote these velocities using the subscript LSR. We use solar velocities from \citet{Schonrich2010}, adopting $(U_\mathrm{\odot LSR}, V_\mathrm{\odot LSR}, W_\mathrm{\odot LSR}) = (11,12,7)$ km/s. The median error in each of components is $3$ km/s, with the error in radial velocity typically dominating.

\subsection{De-biasing the kinematics}\label{Sec:unbiased}

The MEarth sample was selected from a proper motion survey with a lower limit of $0\farcs15$/yr, and are therefore preferentially missing some stars with low tangential velocities. We simulate the stars that we missed due to proper motion limits by drawing velocities and distances from a model thin disk. We consider only the thin disk, because kinematically-hotter stars are a small fraction of the solar neighborhood and less likely to be missed due to proper motion selection. 

We draw $U_\mathrm{LSR}$, $V_\mathrm{LSR}$, and $W_\mathrm{LSR}$ from Gaussian distributions with standard deviations of $35$ km/s, $20$ km/s, and $16$ km/s, respectively \citep{Bensby2003}. We draw distances and positions uniformly in volume. We also tested drawing from the observed distribution of the MEarth sample, and found little difference in the resulting simulated sample, consistent with the conclusions of \citet{Dittmann2014}. We then compute proper motions and apply the $0\farcs15$/yr selection criterion. 

Out to a distance of $25$pc, we find that $11\%$ of stars were missed due to the proper motion limits. Adding in the missing stars, the resulting velocity distributions  for MEarth are similar to the distributions for volume-limited samples of more massive stars \citep[e.g.][]{Holmberg2009}.

\subsection{General kinematic properties of the sample}\label{Sec:kinematics}

\begin{figure*}
\includegraphics[width=0.5\linewidth]{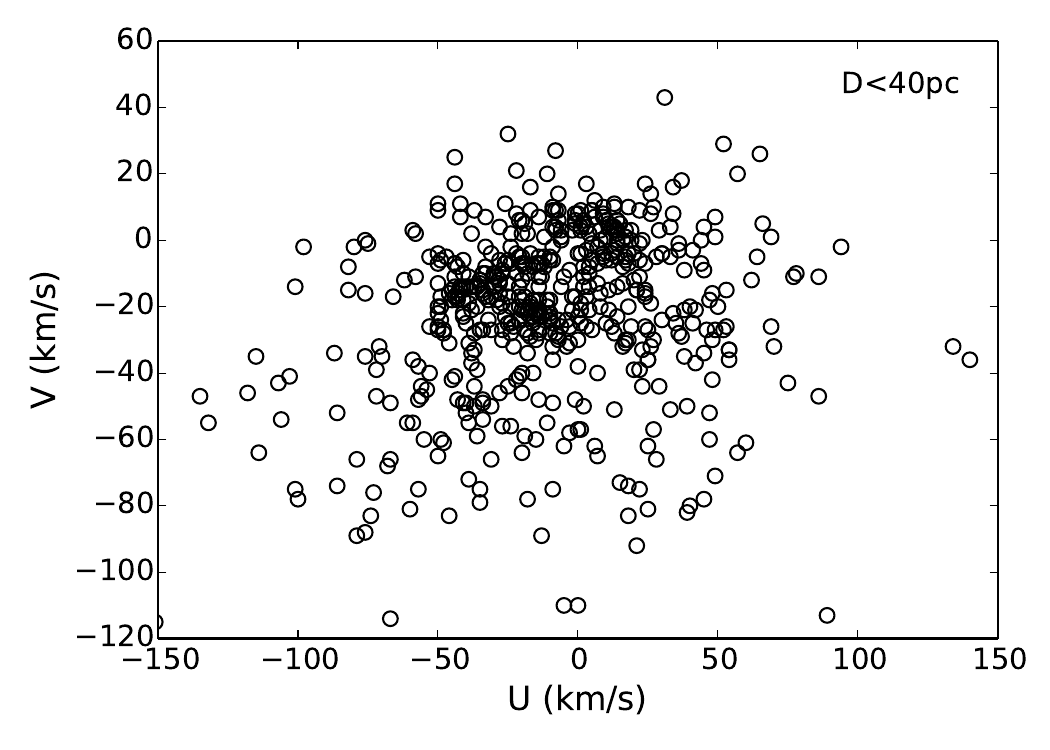}
\includegraphics[width=0.5\linewidth]{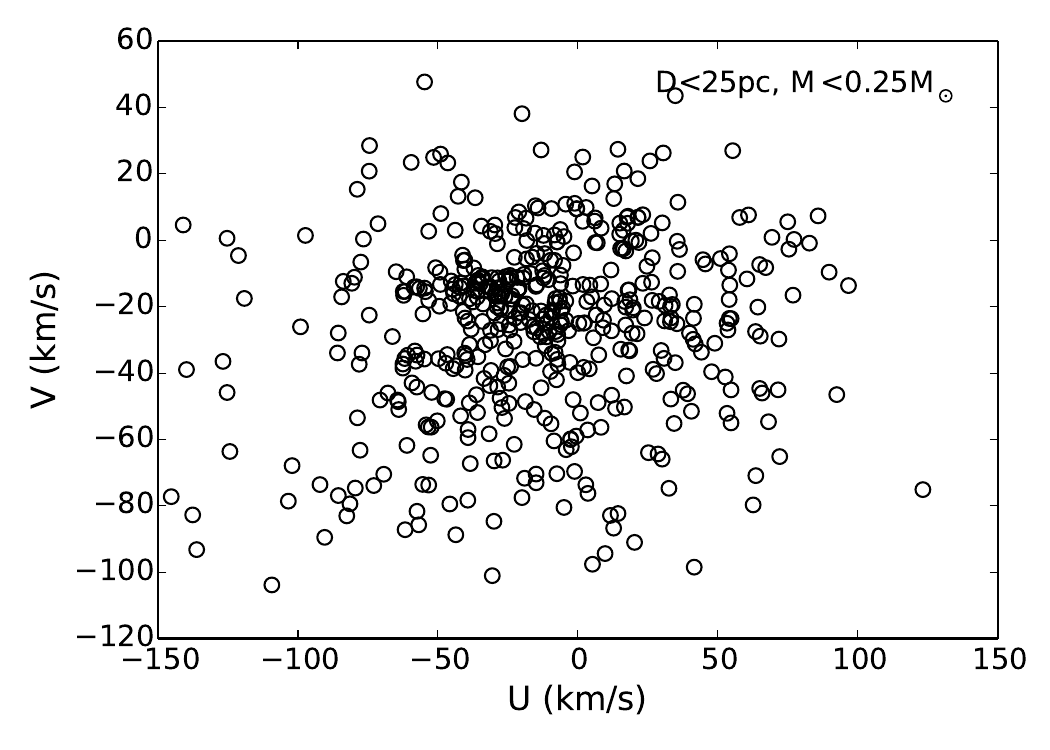}
\caption{$U$ and $V$ velocities (relative to the Sun rather than the local standard of rest) for G and K dwarfs within $40$ pc from the Geneva-Copenhagen Survey (GCS, left) and for mid-to-late M dwarfs within $25$ pc from MEarth (right). The kinematic substructure evident in the GCS is also clearly seen in the nearby M dwarfs; most notable is the Hyades supercluster. The typical error on each component is $3$ km/s.
\label{Fig:galactic}}
\end{figure*}

In Figure \ref{Fig:galactic} we show the $U$ and $V$ velocity components of the Northern MEarth M dwarfs within $25$pc that have estimated masses less than $0.25\ \msun$. We place these limits to mitigate the likelihood of unresolved multiples contaminating the sample. We also show, for comparison, the G and K dwarfs from the Geneva-Copenhagen survey \citep[GCS;][]{Nordstrom2004, Holmberg2009} that are within $40$pc. The kinematic substructures that have been identified for higher mass stars are clearly seen in our mid-to-late M dwarfs as well, most notably the arc at $U\approx-37$ km/s, $V\approx-17$ km/s that has been called the Hyades supercluster, the Hyades stream, and the Hyades moving group \citep{Eggen1958}, not to be confused with the $650$ Myr-old Hyades open cluster.

The Hyades supercluster has similar kinematics to the Hyades and Pleiades open clusters, and at one time the supercluster was proposed to be a stream of stars evaporating from the Hyades open cluster, or at least composed of several coeval groups \citep{Eggen1992, Chereul1998, Chereul1999} though this was not universally agreed upon \citep{Dehnen1998}. However, recent theoretical work shows that spiral structure can dynamically create co-moving groups like the supercluster \citep{DeSimone2004,Quillen2005} and dynamical evolution is thought to be responsible for the larger kinematic structures in the solar neighborhood. For the Hyades supercluster, the dynamical origin of the kinematic association has been demonstrated observationally as well, through analysis of the chemical abundances and the mass function of stars in the proposed supercluster \citep{Famaey2005,Famaey2007,Famaey2008,Bovy2010}.

\subsection{Disk membership}\label{Sec:thickthin}

Stars can be broadly grouped by their kinematics into the thin/young disk, the dynamically-heated thick/old disk, and the even hotter halo population. We assign disk membership using the same method as \citet{Bensby2003}, which takes into account the velocity dispersions in $U_{LSR}$, $V_{LSR}$, and $W_{LSR}$ and the relative number densities of the different stellar populations. The values we use for the velocity dispersions of the thin disk, thick disk, and halo are also from \citet{Bensby2003}. We assume that $89\%$ of the stars in the solar neighborhood are in the thin disk, $10.6\%$ in the thick disk, and $0.4\%$ for the halo \citep{Juric2008}. We do not consider membership in stellar streams.

\begin{figure}
\includegraphics[width=\linewidth]{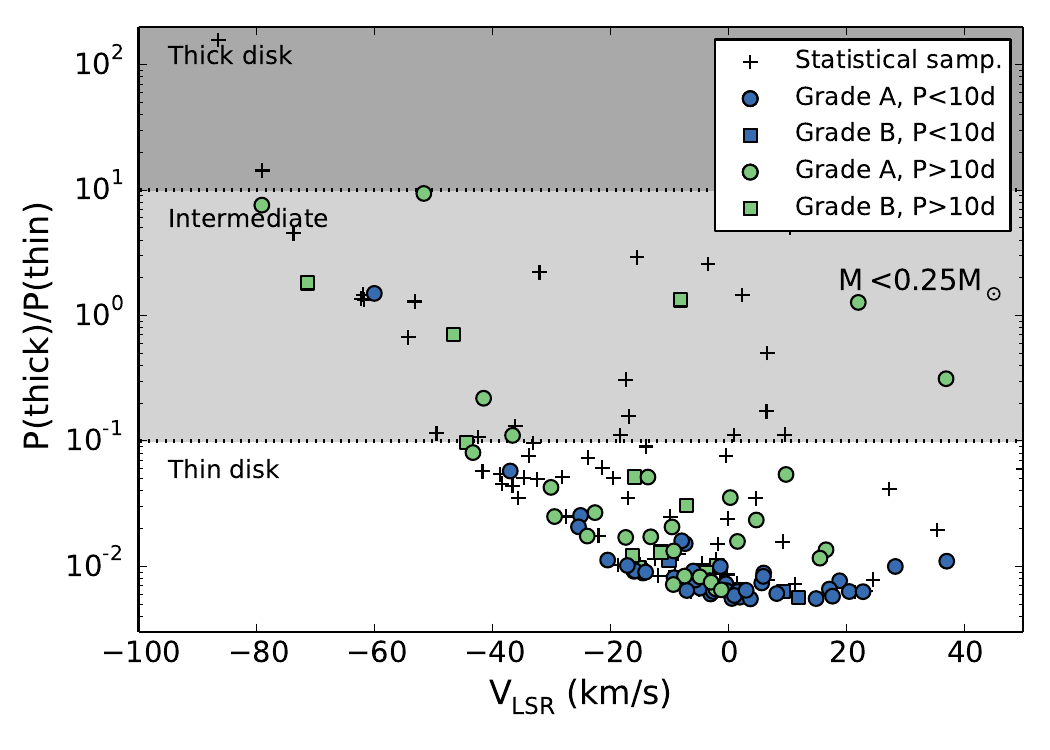}
\caption{Probability ($P$) of an object being in the thick/old disk relative to the probability of it being in the thin/young disk, plotted against $V_\mathrm{LSR}$. $P_\mathrm{thick}/P_\mathrm{thin} =1$ indicates an equal probability of the object being in either disk. Shaded regions denote disk assignments: $P>10$ are thick/old disk members, $0.1<P<10$ are intermediate, and $P<0.1$ are thin/young disk members. Only stars in our statistical sample are shown. Non-detections are plus symbols, grade A rotators are filled circles, and grade B rotators are filled squares. Blue indicates stars with periods shorter than $10$ days, green those with periods longer than $10$ days. \label{Fig:thickthin}}
\end{figure}

In Figure \ref{Fig:thickthin}, we plot the probability of an object being in the thick disk, $P(\mathrm{thick})$, divided by the probability of that object being in the thin disk, $P(\mathrm{thin})$, for the stars in the statistical sample. Out of $163$ stars in the statistical sample that have $UVW$ kinematic information, $23$ ($14\pm3\%$) have $P(\mathrm{thick})>P(\mathrm{thin})$, and $7$ have $P(\mathrm{thick})>10\times P(\mathrm{thin})$. Out of the $87$ grade A and B rotators, $6$ ($7\pm3\%$) have kinematics that potentially place them in the thick/old disk, while none have $P(\mathrm{thick})>10\times P(\mathrm{thin})$. 
Overall, the rotators have kinematics typical of the Solar Neighborhood and are therefore generally members of the thin disk. Rapid rotators, however, are from a dynamically cold population. The p-value for a k-sample Anderson-Darling test \citep{Scholz1987} to check whether the rotators are drawn from the field M dwarf population is $p=10^{-5}$ for rotators with $P<10$ days.

Our results differ from those of \citet{Irwin2011}, who assigned approximately half of the MEarth rotators in their sample to the thick/old disk based on how closely the objects' kinematics matched those expected for each disk. This difference is primarily due to our inclusion of the thin-thick disk normalization.

The fraction of stars with detected periods depends on the kinematic subsample. We divide our statistical sample sample at $P(\mathrm{thick})=0.1\times P(\mathrm{thin})$ to ensure enough stars in the kinematically-older subsample. Our recovery fraction for kinematically-young stars with $M<0.25\msun$ is $58\pm8\%$, while for the kinematically-old stars it is $16\pm8\%$. This may be the result of stars in the kinematically-old subsample generally having longer periods, to which we believe we are less sensitive (\S\ref{Sec:recovery}). Changing spot patterns or variability amplitude could also contribute, though we do not see any such trends amongst the sample of stars for which we do detect rotation periods.

\subsection{Metallicities of the rotators}\label{Sec:metallicity}

\begin{figure}
\includegraphics[width=\linewidth]{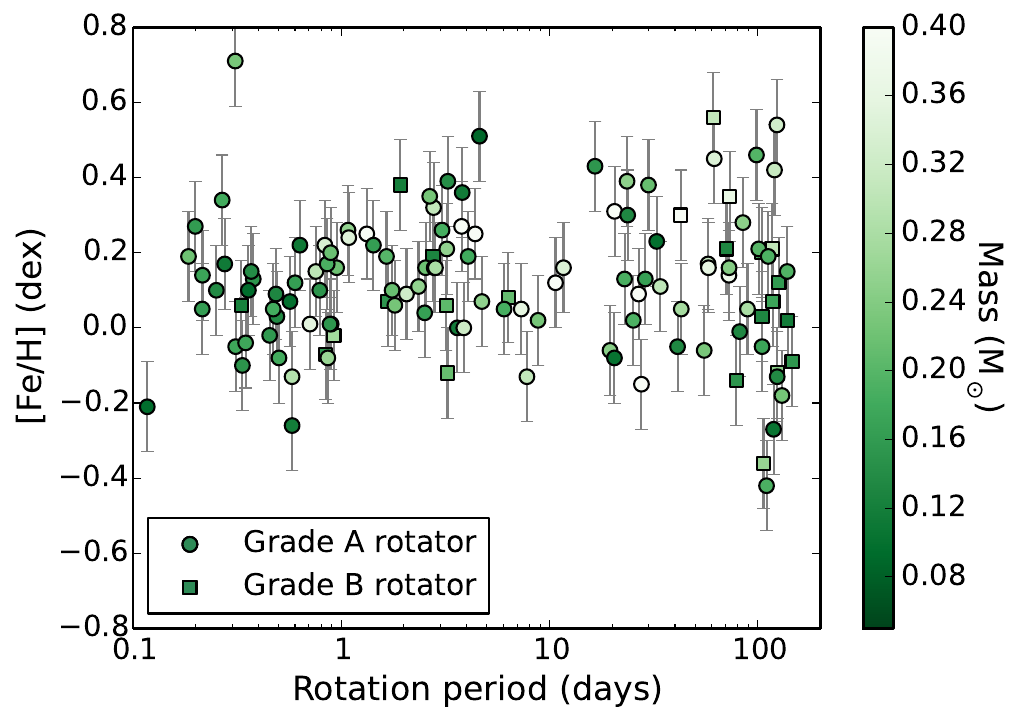}
\caption{$\feh$, estimated from near-infrared spectra, as a function of measured photometric rotation period. The typical error on $\feh$ is $0.12$ dex. The color of the points indicates their stellar mass as estimated from their absolute $K$ magnitudes using our modified version of the \citet{Delfosse2000} relation. We do not see a clear trend between metallicity and rotation.
\label{Fig:feh}}
\end{figure}

\citet{Newton2014, Newton2015} estimated $\feh$ for nearly $450$ MEarth M dwarfs from near-infrared spectra. Figure \ref{Fig:feh}, we show $\feh$ as a function of photometric rotation period. There is not a clear trend with rotation, with a Spearman rank correlation of $0.00\pm0.03$ (see \S\ref{Sec:per-amp}). This is consistent with the interpretation that the rotators are typical Solar Neighborhood stars: within the thin disk, there is no evidence for an age-metallicity relation, and stars may have a range of metallicities \citep[e.g.][]{Nordstrom2004}.

The rotators do not appear significantly more metal-rich than the full sample (Anderson-Darling $p=0.15^{+0.17}_{-0.09}$). We also do not see a correlation between metallicity and period (Spearman correlation coefficient $= -0.02^{+0.03}_{-0.02}$, $p=0.5\pm0.2$), nor between metallicity and amplitude (coefficient $= 0.02^{+0.07}_{-0.08}$, $p=0.5\pm0.2$).

There is one star with unusually large (and unphysical) estimated metallicity of $0.7$ dex: 2MASS J06052936+6049231 (LHS~1817). This star also has large $U$ and $W$ velocities. Two other rotators in our sample have estimated metallicities this high, and both were removed due to known or suspected multiplicity. All three of these objects were also identified as candidate young objects by \citet{Shkolnik2010}, but are not the \emph{only} stars in our sample that are potentially young. Although \citet{Shkolnik2010} have a high-resolution spectrum and did not identify 2MASS J06052936+6049231 as a multiple, its radial velocity ($>100$ km/s) makes it unusual in their sample as well. 

\section{The age-rotation relation}\label{Sec:age}

Because low-mass main sequence stars spin down with time, it is expected that slow rotators are older than their more rapidly-rotating counterparts. While clusters can constrain the rotational evolution at young ages, there are no reliable methods to determine the ages of isolated field M dwarfs -- once they reach the main sequence, their physical properties remain essentially unchanged over a Hubble time. As discussed in the introduction, galactic kinematics provide a means to probe the ages of groups of stars. For example, \citet{Irwin2011} used the total space velocities of $41$ MEarth M dwarfs to classify the stars into the thin/young ($\lesssim3$ Gyr), intermediate ($3-7$ Gyr), and thick/old ($\gtrsim7$ Gyr) disks. \citet{Irwin2011} found that the young disk objects were entirely fast rotators, while the old disk objects were predominantly slow rotators. 

\begin{figure}
\includegraphics[width=\linewidth]{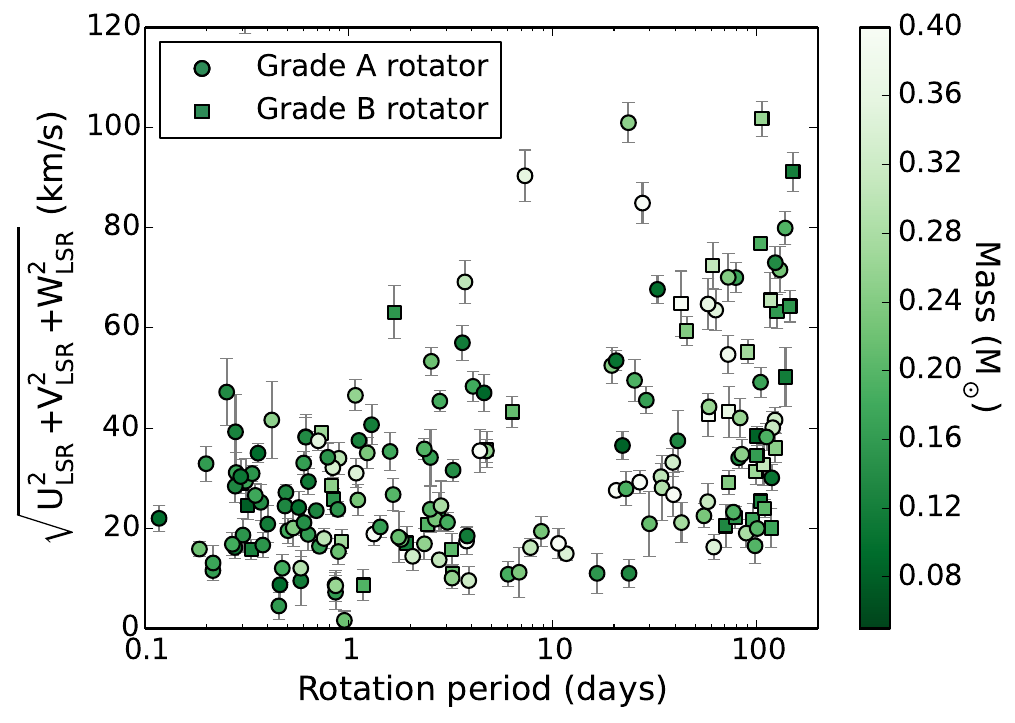}
\caption{Total space velocity as a function of measured photometric rotation period. The color of the points indicates their stellar mass as estimated from their absolute $K$ magnitudes using our modified version of the \citet{Delfosse2000} relation. The velocity dispersion increases with rotation period, as expected if the ages of stars are increasing with rotation period.
\label{Fig:space}}
\end{figure}

We very clearly see the signatures of age-rotation relation in the distribution of total space velocities as a function of photometric rotation period (Figure \ref{Fig:space}). There is an increase in dynamic hotness of the stellar population with rotation period, a trend which spans the entire period range that we probe. This is evidenced in the Spearman rank correlation coefficient for total space velocity and rotation period, which is $0.18\pm0.03$ ($p=0.002^{+0.007}_{-0.002}$). We emphasize, however, that it is the velocity \emph{dispersion} that increases with age. Stars with rotation periods of around a day have space velocities narrowly constrained, as would be expected for a very young stellar population. Stars with rotation periods around 100 days have a wide dispersion in space velocities, as expected for an older stellar population that has been dynamically heated. 

\begin{figure}
\includegraphics[width=\linewidth]{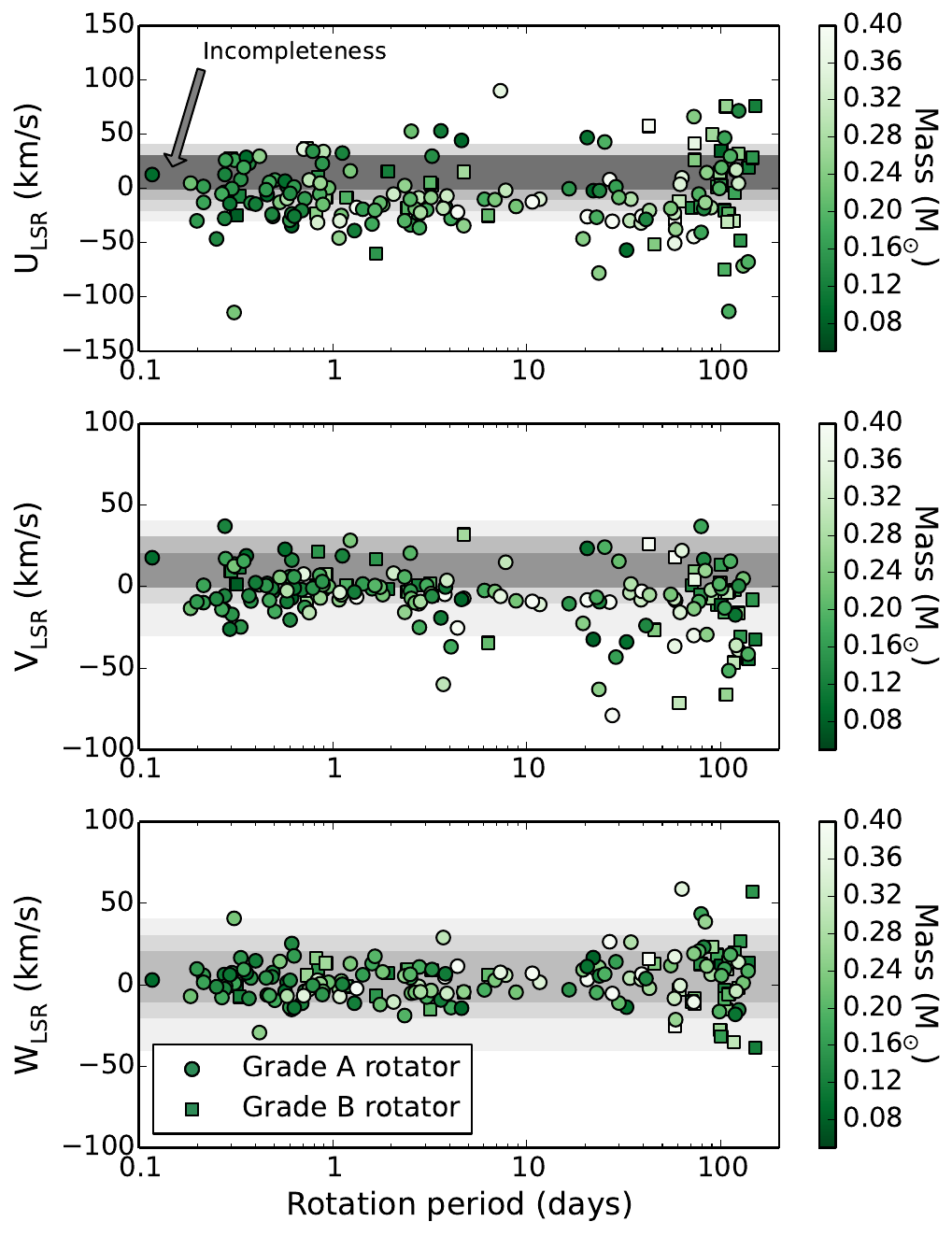}
\caption{Individual components of space velocity as a function of measured photometric rotation period. The color of the points indicates their stellar mass as estimated from their absolute $K$ magnitudes using our modified version of the \citet{Delfosse2000} relation. The velocity dispersion of each component increases with rotation period, as expected if the ages of stars are increasing with rotation period. The $V$ component also becomes increasingly negative (asymmetric drift), which is also a sign of an older stellar population. The gray shaded regions show the precent of stars that were missed in simulations as a result of our selection criteria. We show increments of $5\%$, up to $25\%$, e.g., the darkest gray band shows that at $U$ velocities similar to the Sun, $25\%$ of stars are missing from the MEarth sample. The darkest bands in $V$ and $W$ correspond to $20\%$ and $15\%$, respectively.
\label{Fig:uvw}}
\end{figure}

Decomposing the space velocities into the individual components (Figure \ref{Fig:uvw}), we see the same signatures of aging: the velocity dispersion of each component increases with rotation period. We also see evidence of asymmetric drift: the $V$ velocities become more negative with increasing rotation period. As noted by \citet{West2015}, the magnitude of asymmetric drift is much less than what is seen in more distant, older populations of M dwarfs.

The existence of a relationship between velocity dispersion and age amongst members of the thin disk is well-established \citep[e.g.][]{Wielen1977}, though the exact form of the age-velocity relation is a matter of debate. \citet{Nordstrom2004} and \citet{Holmberg2009}, using data from the GCS fit to a power law, find that velocity dispersion increases smoothly at least to $10$ Gyr. In contrast, \citet{Soubiran2008} find an age-velocity relation with a shallower slope that saturates around $5$ Gyr, using more distant clump giants. They also find significantly higher velocity dispersions, in contrast to studies of the solar neighborhood. \citet{Seabroke2007} found that the data could not constrain whether the age-velocity relationship saturated beyond $5$ Gyr using the data from \citet{Nordstrom2004}. 

Previous kinematic studies of low-mass stars have used the \citet{Wielen1977} age-velocity relations\footnote{Note that \citet{Ofek2009} re-fit the original equations using data from \citet{Nordstrom2004}.} \citep[][]{Schmidt2007, Faherty2009, Reiners2009a, Reiners2010}. We refer the reader to \citet{Reiners2009a} for comments on the usage of the equations. These studies relied on the total velocity dispersion. As discussed by \citet{Seabroke2007}, kinematic substructures such as the Hyades supercluster make the $U$ and $V$ velocity distributions non-Gaussian (see also \S\ref{Sec:kinematics}).
Therefore, we use the $W$ velocity dispersion ($\sigma_W$) for kinematic age assignment. 

In general, we find that the different functional forms and coefficients of age-velocity relations in the literature give consistent results for ages between $1$ and $5$ Gyr. The results diverge for older and younger populations. However, the age-velocity relationship is not appropriate for the youngest stars (as their kinematics are unrelaxed) and not well-constrained at later ages (where it may saturate). Thus, the choice of age-velocity relation is not paramount. 
We adopt the results from \citet[][see Table 6]{Aumer2009}, who modeled the star formation history to arrive at the age-velocity relationship:
\begin{align}
\sigma_W \left( \tau \right) &= \sigma_{W,10} \left( \frac{\tau + \tau_1}{10\; \mathrm{Gyr}+\tau_1} \right)^\beta \\
\sigma_{W,10} &= 23.8\; \mathrm{km/s} \nonumber \\
\tau_1 &= 0.001\; \mathrm{Gyr} \nonumber \\
\beta &= 0.445 \nonumber
\end{align}
The velocity dispersion at $10$ Gyr is described by the parameters $\sigma_{W,10}$ and $\tau_1$, while the exponent $\beta$ characterizes the heating rate.


To apply the age-velocity relation, we first need the dispersion of the $W$ velocity component, $\sigma_W$. To determine the $\sigma_W$ that underlies our data, we take the Bayesian approach of \citet{West2015}, and maximize the posterior probability $p(\sigma_W|D)$, where our data $D$ are our measurements of $W_\mathrm{LSR}$. Using Bayes theorem, the posterior probability is the product of the likelihood, $p(D|\sigma_W)$, and the prior, $p(\sigma_W)$:
\begin{align}
p(\sigma_W|D) \propto p(D|\sigma_W) \times p(\sigma_W)
\end{align}
We use a Jeffreys prior, which is appropriate as an uninformative prior: $p(\sigma_W) \propto 1/\sigma_W$.
The likelihood is the product of the probabilities of obtaining each measurement given the model. The underlying model to which we fit our data is a Gaussian distribution $\mathcal{N}(\mu_W,\sigma_W)$. We use a Cauchy distribution $\mathcal{C}(d_i,\sigma_i)$ to represent our measurement errors. The latter is centered at the measured value and has a standard deviation given by the error ($\sigma_i$) on each datum ($d_i$). This gives:
\begin{equation}
p(D|\sigma_W) = \prod\limits_{i}{V(d_i-\mu_W;\sigma_W,\sigma_i)}
\end{equation}
where $V(W;\sigma_W,\sigma_i)$ is the probability density function (PDF) of a zero-mean Voigt profile, and can be written as the convolution of the PDFs of the Gaussian and Cauchy distributions $N(W;\mu,\sigma)$ and $C(W;\mu,\sigma)$ as:
\begin{equation}
V(W;\sigma_W,\sigma_i) = N(W;0,\sigma_W) \ast C(W;0,\sigma_i)
\end{equation}

The work on higher mass stars on which we base this analysis shows that the average $W$ velocity ($\mu_W$) remains $0$ km/s as the population is dynamically heated. For our sample, $\mu_W$ is close to $0$ km/s as well (Table \ref{Tab:ages}). Due to the small number of objects in our long-period bins, we fix $\mu_W=0$. We choose a Cauchy over a Gaussian distribution to represent measurement errors in order to decrease the sensitivity of the model to outliers (a measurement might be an outlier if, for example, our period measurement is erroneous or if the object is an unidentified multiple). 

We approximate the Voigt PDF following \cite{Thompson1987} as $(1-\eta)\; N(W;0,\Gamma) + \eta\; C(W;0,\Gamma)$. The parameters $\eta$ and $\Gamma$ depend on $\sigma_W$ and $\sigma_i$; they are given in \cite{Thompson1987} and also reproduced in \citet{Ida2000}. The likelihood is then an analytic function that we evaluate at each datum.

We calculate the log of $p(D|\sigma_W)$ for a grid of $\sigma_W$ in step sizes of $0.1$ km/s, and select the $\sigma_W$ that results in the highest posterior probability. We use a bootstrap analysis to estimate errors, sampling with replacement from our data over $100$ iterations.

Our approach should be insensitive to the stars missing from our sample due to MEarth's selection criteria. To test this, we also fit a generative, non-analytic model. In this case, our model is the binned distribution of a random sample of $200000$ stars drawn as discussed in \S\ref{Sec:unbiased} and subject to the MEarth proper motion limit. While we are interested only in $\sigma_W$, the velocity dispersions of the other velocity components are not independent so we fix $\sigma_U$ and $\sigma_V$ in the ratios of the thin disk.  After drawing our sample, we apply the $0\farcs15/\mathrm{yr}$ proper motion limit. As we demonstrated in \S\ref{Sec:unbiased}, the selection of our sample using proper motions causes our sample to be missing $11\%$ of the stars, but the fraction of stars missing will be larger when the velocity dispersion is smaller. We then convolve the resulting PDF with a Cauchy distribution to account for the errors on each datum. The results from this approach are similar to those from the simpler method we adopt.

We divide the sample into bins in period, $P<1$ day,  $1<P<10$ days, $10<P<70$ days, and $P>70$ days, considering only objects with $0.1<M<0.25\ \msun$. Considering all rotators (both grade A and B), we infer mean ages of (0.5, 0.7, 0.6, 5.1) Gyr in these bins, respectively. We arrive at similar results using $2.3<MEarth-K<3.3$ to select our low-mass sample.

We also apply the \citet{Wielen1977} relation as described in \citet{Reiners2009a}. 2MASS J06052936+6049231, the rapidly rotating star with a very negative $U$ velocity seen in Fig. \ref{Fig:uvw}, strongly affected the results and was excluded. The ages we infer considering all rotators for the four bins defined above are (0.7, 1.7, 3.1, 5.4) Gyr. These ages \emph{are} affected by our proper motion bias since they rely on calculating the dispersions of the observed sample.

We present our results in Table \ref{Tab:ages}. Within the errors, stars with $P<10$ days have $\sigma_W\lesssim10\ \mathrm{km/s}$, implying ages of $<1$ Gyr according to our chosen age-velocity relation. This is younger than the youngest bin used in the calibration, and the distribution of velocities in the GCS is fairly constant from $1$ to $2$ Gyr. We therefore assign this population of stars mean ages of $\lesssim2$ Gyr. Our results for $10<P<70$ days are not robust: there are relatively few stars at these periods, there is a strong dependence on the upper period boundary, and the total space velocities indicate an older population that the $W$ component alone. For the longest-period rotators, with $P>70$ days, we adopt a mean age of $5^{+4}_{-2}$ Gyr.

The velocity dispersions we determine are slightly lower than those from \citet{West2015}, which is also based on the sample of MEarth rotators, and therefore we obtain slightly younger ages than one would infer from their work. Due to the mass dependence of rotational evolution (see \S\ref{Sec:mass-per}), we restricted the range of masses used in this analysis. If we include rotators regardless of mass, we arrive at slightly larger velocity dispersions. Our work also includes a compilation of other published radial velocities, so more precise measurements are available for some objects, and we have made efforts to remove possible multiples, which may have velocities or periods uncharacteristic of otherwise similar stars. We also only use stars with trigonometric distance measurements.

\begin{deluxetable*}{l r r r r r}
\tablecaption{\label{Tab:ages}Velocity dispersions and ages for stars with detected rotation periods}
\tablecolumns{6}
\tablehead{ 
	\colhead{Period bin} & 
	\colhead{N stars} &
	\colhead{Mean P} &
	\colhead{Mean $W_\mathrm{LSR}$} &
	\colhead{$\sigma_W$} &
	\colhead{Est. age} 
	\\
	\colhead{(days)} &
	 &
	\colhead{(days)} &
	\colhead{(km/s)} &
	\colhead{(km/s)} &
	\colhead{(Gyr)}
	}
\startdata
\sidehead{Grade A}
\hline \\ 
$0<P<1$ & $39$ & $ 0.5$ & $ 3$ & $ 6.0^{+ 1.8}_{- 1.0}$ & $ 0.5^{+ 0.4}_{- 0.2}$ \\ 
$1<P<10$ & $23$ & $ 2.9$ & $ 0$ & $ 7.4^{+ 1.8}_{- 1.8}$ & $ 0.7^{+ 0.5}_{- 0.3}$ \\ 
$10<P<70$ & $10$ & $28.3$ & $ 4$ & $ 6.5^{+ 1.6}_{- 1.5}$ & $\cdots$ \\ 
$P>70$ & $14$ & $102.4$ & $ 9$ & $16.7^{+ 5.3}_{- 4.5}$ & $ 4.5^{+ 3.9}_{- 2.3}$ \\ 
\sidehead{Grade A+B}
\hline \\ 
$0<P<1$ & $43$ & $ 0.5$ & $ 3$ & $ 6.3^{+ 1.6}_{- 1.3}$ & $ 0.5^{+ 0.3}_{- 0.2}$ \\ 
$1<P<10$ & $31$ & $ 2.9$ & $ 1$ & $ 7.3^{+ 1.4}_{- 1.4}$ & $ 0.7^{+ 0.3}_{- 0.3}$ \\ 
$10<P<70$ & $11$ & $29.9$ & $ 5$ & $ 6.9^{+ 1.6}_{- 1.8}$ & $\cdots$ \\ 
$P>70$ & $28$ & $106.2$ & $ 6$ & $17.7^{+ 5.4}_{- 4.7}$ & $ 5.1^{+ 4.2}_{- 2.6}$ 
\enddata
\end{deluxetable*}

\section{The mass-period relation}\label{Sec:mass-per}

Rotation is found to be strongly mass-dependent in young open clusters, with the lowest-mass stars reaching the fastest rotation rates and maintaining rapid rotation for longer.
Rotational evolution at field ages is also mass dependent. Lower-mass stars spin down more slowly than higher-mass stars on the main sequence, but eventually reach longer rotation periods. This mass dependence in the upper envelope of rotation periods is clearly seen in \citet[][mid and late M dwarfs from MEarth]{Irwin2011} and \citet[][early M dwarfs from \kepler]{McQuillan2013,McQuillan2014}.

We draw on the large sample of M dwarf photometric rotation periods measured from the \kepler\ survey by \citet{McQuillan2013} to explore the rotation period distribution across the M dwarf spectral class. 
We downloaded additional data on these stars from the Kepler Input Catalog \citep[][KIC]{Brown2011} from the MAST.

\begin{figure*}
\includegraphics[width=0.5\linewidth]{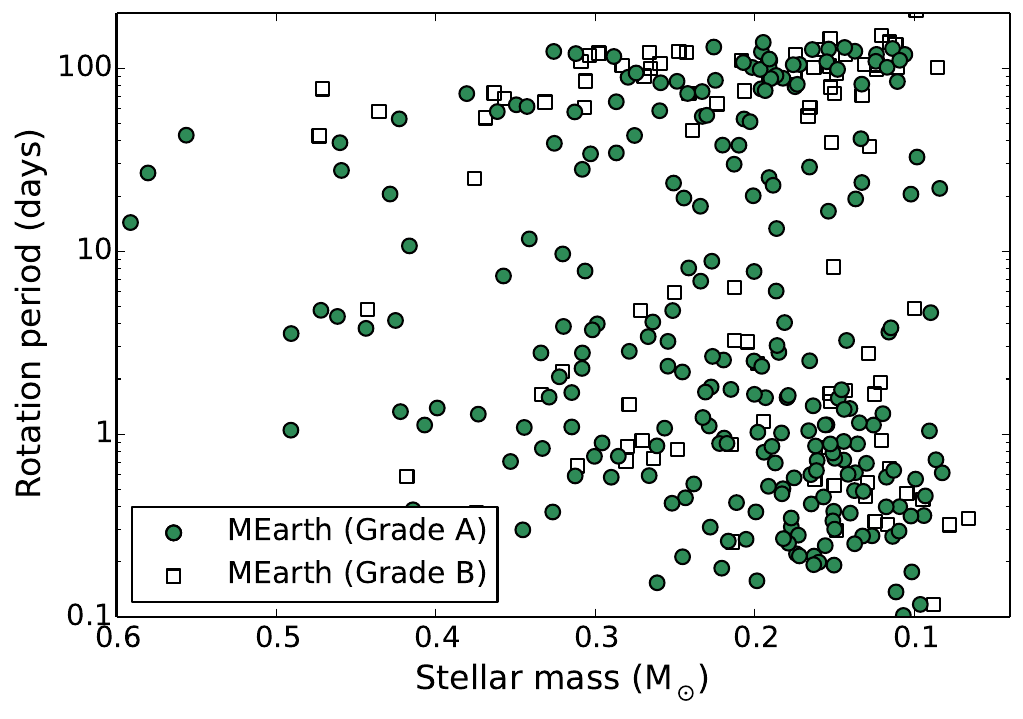}
\includegraphics[width=0.5\linewidth]{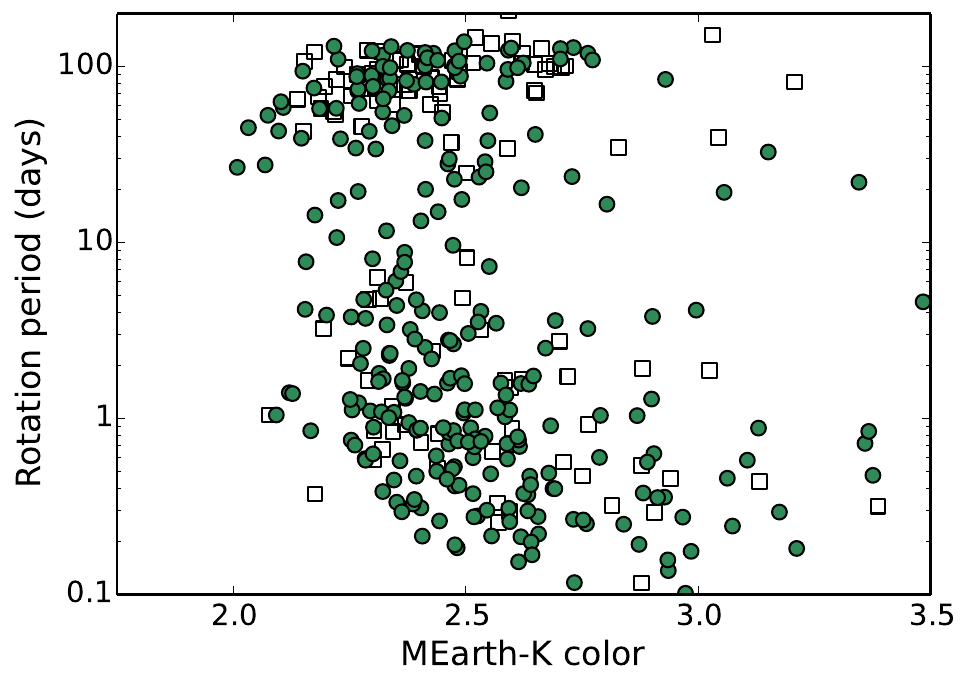}
\includegraphics[width=0.5\linewidth]{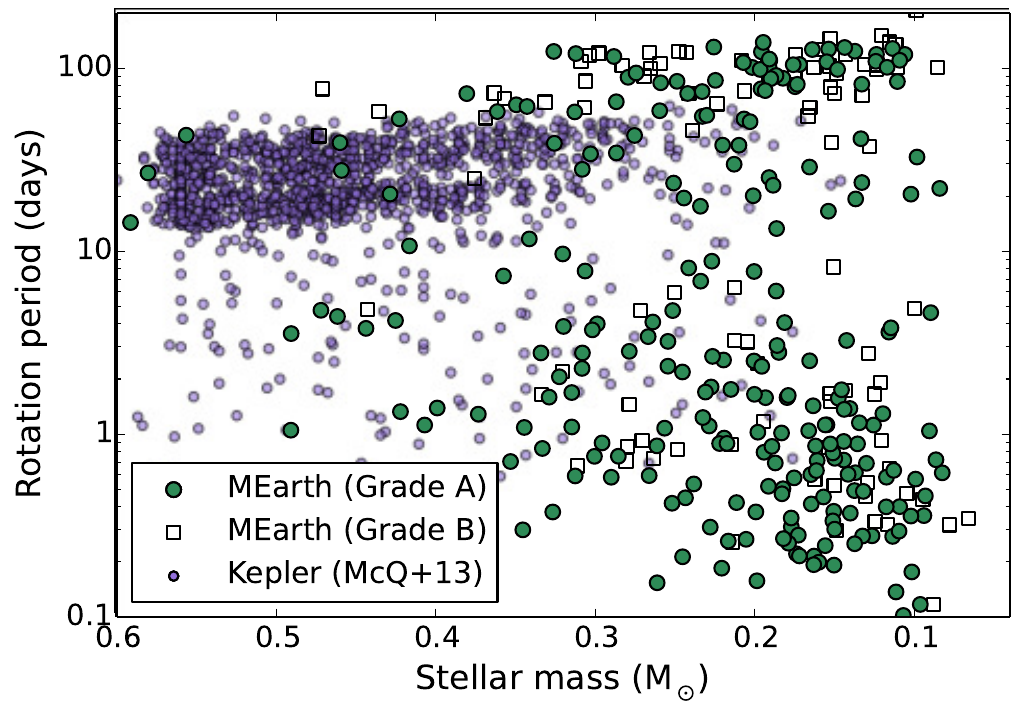}
\includegraphics[width=0.5\linewidth]{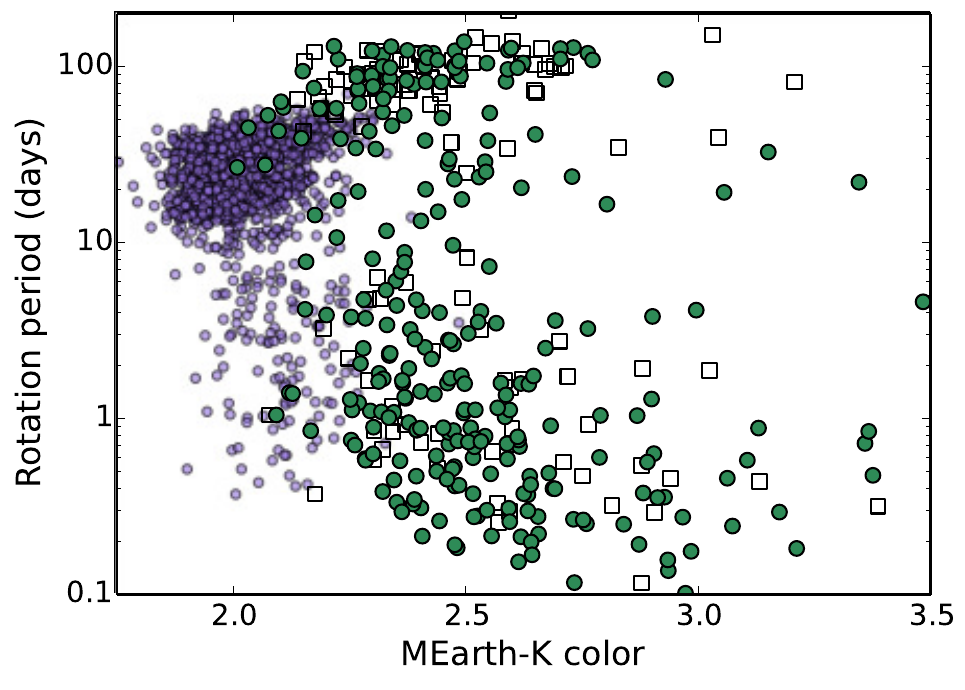}
\caption{Period versus stellar mass (left panels) and versus color (right panels). Only MEarth rotators from this work are shown in the upper panels, while in the lower panels we include \kepler\ rotators from \citet{McQuillan2013} and MEarth rotators. The masses of the \kepler\ stars are estimated from broadband colors and stellar models, while the masses for the MEarth stars are estimated from absolute K magnitudes. We use a linear combination of $i$ and $z$ magnitudes to estimate the magnitude of the \kepler\ stars in the MEarth bandpass. We have removed known and suspected multiples from these plots, but the MEarth stars with $M>0.3\msun$ are more likely to be unresolved multiples due to MEarth's selection criteria. We see evidence for both a lower and upper envelope on the detected periods in the MEarth sample, and a lack of stars with intermediate rotation periods.
\label{Fig:mass}}
\end{figure*}

Absolute magnitudes provide the best way to estimate masses for single M dwarfs, and we use parallaxes to obtain absolute \K\ magnitudes for the MEarth rotators as described in \S \ref{Sec:method_period}. Parallaxes are not available for the majority of M dwarfs targeted by \kepler, so we instead use masses estimated by \citet{Dressing2013}, who matched broadband photometry to Dartmouth stellar models.\footnote{The requisite multicolor photometry for this method is not available for the brighter MEarth M dwarfs} Because the masses for MEarth and \kepler\ are determined using different methods, there may be an offset between the two mass scales. MEarth stars were selected to have $R<0.33\rsun$, so any star with a mass greater than about $0.3\msun$ is brighter than expected and more likely to be an unresolved multiple. Figure \ref{Fig:mass} plots photometric period versus estimated mass across the entire M spectral class. 

Because we are interested in the mass-period relation, it is important that we have a uniform basis on which to compare the MEarth and \kepler\ samples. We therefore turn to photometric colors. The only reliable optical magnitude available for all MEarth M dwarfs is the apparent magnitude in the MEarth bandpass, which was calibrated by \citet{Dittmann2015}, so we use $\mearth-K$ colors. The MEarth bandpass comprises most of $i$ and all of $z$, so it is possible to estimate $\mearth$ magnitudes for the \kepler\ stars from the KIC photometry with reasonable accuracy. We use an empirical relation derived from presently-unpublished observations we have obtained of a number of MEarth targets in the SDSS filters:
\begin{align}
\mearth = (i + 2\times z) / 3 - 0.20
\end{align}
This relation has a scatter of about $5\%$. Masses from \citet{Dressing2013} account for reddening; our color transformation does not. Figure \ref{Fig:mass} also plots photometric rotation period versus color.

The long rotation periods we find for the mid M dwarfs are consistent with the previous MEarth results from \citet{Irwin2011}, who found that the lower the mass of the star, the longer the period to which it spins down.
It is challenging to infer the shape of the upper period envelope due to the lack of overlap between the stellar populations probed by \kepler\ and by MEarth. \citet{McQuillan2013} did not detect rotation periods longer than $70$ days in any of their objects, although periods up to $155$ days were searched.\footnote{In \citet{McQuillan2014} they place an upper limit of $70$ days on the periods searched, but they do not in this earlier work.} However, it is possible that \kepler's systematics, particularly differences between \kepler's data quarters, affect the recovery of longer rotation periods.

The lower envelope of the period distribution (shortest period seen at a given mass) is also mass-dependent, with the most rapid rotators having shorter periods at lower masses, particularly below the full convection limit. This feature is also seen in the Hyades and Praesepe open clusters \citep[e.g.][]{Agueros2011} and in \vsini\ studies of field late M and brown dwarfs \citep{Mohanty2003, Jenkins2009}.

We find stars with intermediate rotation periods less often than more slowly-rotating stars, and that the gap between ``slow'' and ``fast'' rotators increases with decreasing mass. 
Using $41$ light curves from the 2008-2010 seasons of MEarth data, we showed in \citet{Irwin2011} that completeness was independent of rotation period for $P<100$ days. This implies that the gap is astrophysical.
Returning to Figure \ref{Fig:distribution}, which shows the distribution of periods for stars in our statistical sample, the lack of stars at intermediate periods is clear in the grade A and grade B rotators, as well as in the candidate periods for the ``possible'' detections.
A similar gap has been seen in activity studies of M dwarfs \citep{Herbst1989, Gizis2002,Cardini2007, Browning2010}.

We suggest that the most likely explanation for the gap is that these mid M dwarfs spin down rapidly from ``fast'' to ``slow'' rotation rates. Under this hypothesis, M dwarfs spend comparatively little time at intermediate rotation periods, making it unlikely we would catch them there by chance in a field population with a wide range of ages.

\section{Summary}

We have searched for photometric rotation periods in every star that has been observed by the Northern MEarth transit survey. The rotation periods and ratings we present here supersede those reported previously in \citet{Irwin2011} and \cite{West2015}.
The comparison of our rotation periods to other photometric periods from the literature and to \vsini\ measurements lends support to the periods we have detected, although we refer the reader to \S\ref{Sec:comparison} for further discussion.

The rotation periods we detect range from $0.1$ days to $140$ days. Due to our requirement that the photometric modulation be repeated, we expect that we may not be able to detect periods longer than $140$ days (about half of the longest possible observing season), so this limit may simply reflect the longest period to which we are sensitive. For fully-convective stars with detected rotation periods, the amplitude of variability is independent of the rotation period, and we find no correlation between metallicity and rotation period or amplitude. Amongst rapid rotators, we find an abundance of stable, sinusoidal modulations.  Our recovery rate in the subset of best-observed stars is $47\pm5\%$, and is higher for kinematically-young stars than it is for kinematically-old stars.

We used the variety of data that our team has collected on these stars to probe the Galactic kinematics of mid M stars in the Solar Neighborhood and of rotators in particular. Accounting for the selection criteria for the MEarth sample, we found that the nearby mid M dwarfs have kinematics consistent with those of higher-mass stars. We found evidence of the substructure seen in the kinematics of higher-mass stars amongst the M dwarfs as well, in particular the dynamically-created Hyades supercluster. These substructures, which most strongly affect the $U$ and $V$ components of the space velocities, are important to consider when drawing conclusions about the kinematics of local groups of stars.

There is clear evidence for a rotation-age relation in all three velocity components. Using the dispersion in the $W$ velocity component and established age-velocity relationships, we estimated the mean ages for different populations of rotators. Considering M dwarfs with $0.1<M<0.25\ \msun$, we found that stars with rotation periods less than $10$ days are on average less than $2$ Gyr old, while the slowest rotators we estimate to have an average age of $5^{+4}_{-2}$ Gyr. We find that most rotators are likely members of the thin/young disk. 

The mass-period relationship, as traced by the MEarth and \kepler\ M dwarfs, confirms that mid M dwarfs spin down to longer periods than earlier M dwarfs. The fastest rotation periods we found amongst the field stars decrease with decreasing mass. We also see a lack of stars with intermediate rotation periods.

\section{Conclusions}

Our results are consistent with a scenario in which mid-type M dwarfs maintain rapid rotation (and enhanced magnetic activity) for the first several billion years of their life. At the age of the Hyades and Praesepe, M dwarfs have a range of rotation rates, with the latest-type M dwarfs having periods of $<1$ day \citep{Scholz2007}. Our field M dwarfs with periods $<10$ days are likely not much older than these clusters, given their low velocity dispersion. These stars do not appear to have converged to the same narrow mass-period relationship on which more massive stars are found. Convergence erases the dependence of rotation periods on the initial conditions, which is a prerequisite for gyrochronology.

We see an increase in the dispersion of total space velocity for increasingly longer periods, demonstrating that gyrochronology is potentially feasible for mid M dwarfs at old ages and rotation periods of about 100 days, if convergence can be established.
Our current sample only allows us to to determine that the mean age of those M dwarfs with $P>70$ days is about $5$ Gyr. This may represent a sample that is continuing to spin down slowly, and for which rotation period increases with age. More precise constraints on the age-rotation relation at long periods are required.

We have demonstrated that Galactic kinematics is useful tool for studying the age-rotation relation, and with a larger sample of stars will provide further constraints. However, the use of kinematics is limited by our understanding of the age-velocity relation, which at present is best calibrated from $1$ to $5$ Gyr. Due to the population-level approach our analysis requires, kinematics may not be able to establish whether the rotation periods of mid M stars converge. M dwarfs in multiple systems with stars of determined ages, such as white dwarfs, provide another promising avenue \citep[e.g.][]{Morgan2012,Rebassa-Mansergas2013,Dhital2013}. Observations of M dwarfs in older clusters, while potentially quite useful, are technically challenging due to the greater distances at which these clusters are found, the relative faintness of the M dwarf members, and the need to establish cluster membership, but may become feasible with future observational advances.

\citet{McQuillan2013}, using \kepler\ photometry, found that early-type M dwarfs with periods less than $10$ days had high amplitudes of variability and stable spot patterns. They postulated that these objects were binaries, which could also explain why no candidate planets have been found around them. Extrapolation of the period distribution of the rapidly-rotating mid-type M dwarfs from MEarth indicates that the young, early-type field M dwarfs should have periods of $1-10$ days. The stability that \citet{McQuillan2013} saw is also reminiscent of the well-behaved sinusoids we see in rapidly-rotating, lower-mass M dwarfs. This suggests that young field stars could be a substantial component of the rapidly-rotating \kepler\ M dwarf sample.

The relative lack of field mid M dwarfs with intermediate rotation periods -- between about $10$ and $70$ days -- supports the suggestion of \citet{Irwin2011} that spin-down occurs rapidly. The gap in periods is similar to that seen in the distribution of magnetic activity levels, and may be the result of the same underlying physical mechanism.
The rapid evolution may occur when the stars reaches a critical condition, which could be a certain rotation rate or magnetic flux. It could also relate to a change in magnetic field topology, which more effectively couples the stellar wind and magnetic field \cite[see e.g.][]{Garraffo2015}. The mass-period relation shown in Figure \ref{Fig:mass} suggests that the critical condition may be mass-dependent, as the gap appears to narrow at earlier spectral types. 
Using the mean age of our rapid and slow rotators as the lower and upper bounds, we suggest that this occurs between $2$ and $5$ Gyr. 

The active lifetime of mid and late M dwarfs is plausibly associated with the rapid evolution we discuss above. \citet{West2015} found the fraction of active stars (as traced through H$\alpha$ emission) decreases significantly for the longest-period rotators in the MEarth sample. \citet{West2008} determined that activity lifetime is about $5$ Gyr for M4V stars, and $7$ Gyr for M5V stars (a large jump in active lifetime is seen between M3V and M5V, which these authors associate with the fully-convective boundary). Our work implies a somewhat shorter active lifetime, but this may be the result of the different age-velocity relationship used by \citet{West2008}, which assumes a steeper power law and no saturation.

Stars with rotation periods of around $100$ days are not generally found to be magnetically active \citep{West2015}. Nevertheless, many slowly-rotating mid-to-late M dwarfs show variability amplitudes of half a percent or more, implying that they have maintained strong enough magnetic fields to produce the requisite spot contrasts. The lack of correlation between rotation period and amplitude for these stars indicates that the spot contrast is not changing significantly, even while they undergo substantial spin-down.

We are collecting additional $H\alpha$ measurements and radial velocities to further improve our understanding of the connection of magnetic activity and kinematics to rotation, and using the MEarth-South data to search for new rotators amongst the nearby M dwarfs in the Sorthern hemisphere. Our goal is to further constrain the age-rotation-activity relation, particularly at intermediate and long periods.

\acknowledgments The MEarth project acknowledges funding from the National Science Foundation under grants AST-0807690, AST-1109468, and AST-1004488 (Alan T. Waterman Award) and the David and Lucile Packard Foundation Fellowship for Science and Engineering. This publication was made possible through the support of a grant from the John Templeton Foundation. The opinions expressed here are those of the authors and do not necessarily reflect the views of the John Templeton Foundation. ERN was supported by the NSF Graduate Research Fellowship, and ZKB-T by the MIT Torres Fellowship for Exoplanet Research. AAW acknowledges the support of NSF grants AST-1109273 and AST-1255568 and the Research Corporation for Science Advancement's Cottrell Scholarship.

This research has made use of data products from the Two Micron All Sky Survey, which is a joint project of the University of Massachusetts and the Infrared Processing and Analysis Center / California Institute of Technology, funded by NASA and the NSF; the Sloan Digital Sky Survey (SDSS); NASA's Astrophysics Data System (ADS); and the SIMBAD database and VizieR catalog access tool, at CDS, Strasbourg, France.

\clearpage

\end{document}